\documentclass[snbasic]{sn-jnl}
\usepackage{multirow}
\usepackage[numbers, sort&compress]{natbib}  
\usepackage{enumerate}

\begin{document}

\title[Utilizing Source Code Syntax Patterns to Detect BIC using ML Model]{Utilizing Source Code Syntax Patterns to Detect Bug Inducing Commits using Machine Learning Models}


\author*[1]{\fnm{Md} \sur{Nadim}}\email{mdn769@usask.ca}

\author[2]{\fnm{Banani} \sur{Roy}}\email{banani.roy@usask.ca}

\affil*[1]{\orgdiv{Ph.D. Student, Software Research Lab (SRLab), Department of Computer Science}, \orgname{USASK}, \orgaddress{\city{Saskatoon}, \state{Saskatchewan}, \country{Canada}}}

\affil[2]{\orgdiv{Assistant Professor, Department of Computer Science}, \orgname{USASK}, \orgaddress{\city{Saskatoon}, \state{Saskatchewan}, \country{Canada}}}


\abstract{Detecting Bug Inducing Commit (BIC) or Just in Time (JIT) defect prediction using Machine Learning (ML) based models requires tabulated feature values extracted from the source code or historical maintenance data of a software system. Existing studies have utilized meta-data from source code repositories (we named them GitHub Statistics or GS), n-gram-based source code text processing, and developer’s information (e.g., the experience of a developer) as the feature values in ML-based bug detection models. However, these feature values do not represent the source code syntax styles or patterns that a developer might prefer over available valid alternatives provided by programming languages. This investigation proposed a method to extract features from its source code syntax patterns to represent software commits and investigate whether they are helpful in detecting bug proneness in software systems. We utilize six manually and two automatically labeled datasets from eight open-source software projects written in Java, C++, and Python programming languages. Our datasets contain 642 manually labeled and 4,014 automatically labeled buggy and non-buggy commits from six and two subject systems, respectively. The subject systems contain a diverse number of revisions, and they are from various application domains. Our investigation shows the inclusion of the proposed features increases the performance of detecting buggy and non-buggy software commits using five different machine learning classification models. Our proposed features also perform better in detecting buggy commits using the Deep Belief Network generated features and classification model. This investigation also implemented a state-of-the-art tool to compare the explainability of predicted buggy commits using our proposed and traditional features and found that our proposed features provide better reasoning about buggy commit detection compared to the traditional features. The continuation of this study can lead us to enhance software effectiveness by identifying, minimizing, and fixing software bugs during its maintenance and evolution.



}




\maketitle

\section{Introduction}
\label{sec:introduction}
There are several studies on detecting buggy commits or Just in Time (JIT) defect predictions \cite{Kim:Classify:Clean:Buggy, Rel:Kim:2007:PFC:1248820.1248881, Rel:RePEc:ids:ijrsaf:v:7:y:2013:i:1:p:17-31, Rel:2013:Shivaji:Reducing:Feature:Bug:Prediction, Rel:2015:Yang:Deep:JIT:Defect:Prediction, Rel:Kim:2006:AIB:1169218.1169308, Sliwerski-2005-induce-fixes-journal, Rel:Wen:2016:LLB:2970276.2970359}; most of these studies either used statistics extracted from their respective GitHub repositories as features \cite{2019:Borg:SZZUnleashed, CommitGuruRosen} or features extracted from  n-gram \cite{cavnar1994n-gram} based source code processing \cite{ReducingFeatures-BIC}. These studies can provide assumptions about whether a commit is bug-inducing or clean, but they cannot provide any specific idea/ reasoning about how such bugs have been induced or what could be done to fix those induced bugs. Existing Studies \cite{Kamei:2013:LES:2498737.2498844, Kim:Classify:Clean:Buggy, Sliwerski-2005-induce-fixes-journal, Yin:2011:FBB:2025113.2025121, Has-bug-really-fixed} made it clear that failure in properly doing the maintenance activity of a software system is one of the most common reasons for introducing bug(s). A developer who is responsible for doing such maintenance activity performs coding in some unique syntax patterns or styles. 

It is possible to have two different structures and code complexity of the code fragment written to solve the same problem if two other programmers write those. This study will practically verify the effect of such a difference in coding syntax patterns in detecting bug-inducing  commits utilizing ML-based detection models. \citet{CLEVER-JIT} reported the use of a clone detection tool named Nicad \cite{5970189Nicad} to determine the similarity of a new buggy code fragment to the code fragment(s) where a bug had already been fixed previously for suggesting the possible bug fix to the developer. There are a few more similar studies \cite{BugFix-Debugging-Helper, APR-Patch-Code, MiningBugFix-APR, FixingIngredientsExists} that investigated generating bug-fix patterns based on the existing commits patches and identical code fragments. Suppose we can identify a specific set of coding syntax styles/ patterns responsible for inducing bugs/ inconsistencies in software systems. We can also determine the similarity of those patterns to the new developers' code and provide general guidelines to the developers to avoid such patterns or structures in the codebase. More generalized (e.g., commercial and open-source software systems of different sizes and application domains) recommendations about risky commits and their possible fixes can be suggested to software developers using the dangerous patterns identified from the commits, which are most likely to induce bugs in software systems. 

\citet{NaturalPreference} show that the source code of a software system shows more repetitiveness compared to the natural language text, and developers prefer some specific implementation of source code statements over other valid alternatives. For example, let us consider a simple $if$ statement is written in the C or Java programming language $if(x<=y)$ or $if(x<(y+1))$; both the code fragments have the same meaning but are written in two different syntax patterns/ style. Different syntax styles or patterns imply differences in some basic coding elements, such as the number of identifiers, the implementation of conditions, nesting levels of different statements, and the use of inter-dependent statements. The preferences of software developers also may affect the ordering of programming tokens, such as the following two valid alternatives, a) i = i + 1, b) i = 1 + i. Although both alternative statements have the same meaning when it is compiled on a computer, they might be perceived in diverse ways when a software developer is working in that software system later. The difference in such perceptions and preferences by different programmers/ developers of a software system might add a different level of complexity during software maintenance and evaluations. A similar scenario may also happen when different developers are coding complex or larger functionalities.This paper show investigations to verify whether such a difference in developers’ coding syntax patterns also makes a difference in complexity level while changing or fixing the identified issues in those code fragments by detecting the buggy code commits using machine learning and deep learning classification models. Our investigation utilizes five different ML-based classification models and compares the performance improvement of BIC detection using GitHub statistics and code syntax pattern based features. 

Our goal in this study is to extract and evaluate developers' coding syntax patterns-based features in detecting bug-inducing commits (BICs) using machine learning (ML) classification models. The analysis of these features will help us identify the distinctive coding syntax patterns from bug-inducing  and clean commits. This identification of these patterns can provide a meaningful idea or reasoning to the software developer about the induced bug in the software system. \citet{BoostingAPR-BIC} published their preliminary fixing strategy devised by the patterns of Bug Inducing Commits, and their study reported automatic fix/ repair of eight new bugs, which the state-of-the-art techniques could not repair. They provided importance on analyzing BIC patterns rather than analyzing how bugs are being fixed to boost the automatic program repair (APR) \cite{APR-1, Tbar-fix-pattern, APR-Fixing-Contracts, APR-Human-Patch, BetterCodeSearchAPR, ICSE2020-DLFix:APR} techniques. Our study to evaluate and identify the developers' coding syntax patterns responsible for inducing bugs in software systems could also contribute to finding relevant bug fixing patterns. 


\begin{table}
\centering
\caption{\textsc{Dataset Summary-I (Manually Labeled \cite{2019:FSE:Wen:EEC:3338906.3338962})}}
\label{tab:data-summary-i}
\begin{tabular}{|l|l|c|c|c|} 
\hline
\multicolumn{1}{|c|}{\begin{tabular}[c]{@{}c@{}}\textbf{Subject}\\\textbf{ Systems}\end{tabular}} & \multicolumn{1}{c|}{\begin{tabular}[c]{@{}c@{}}\textbf{Application }\\\textbf{Domains}\end{tabular}} & \begin{tabular}[c]{@{}c@{}}\textbf{BIC}\\\textbf{ (L = 1)}\end{tabular} & \begin{tabular}[c]{@{}c@{}}\textbf{CC }\\\textbf{ (L = 0)}\end{tabular} & \textbf{BIC+CC}          \\ 
\hline
\textbf{Accumulo}                                                                                 & \begin{tabular}[c]{@{}l@{}}Data Storage \\and Retrieval\end{tabular}                                 & 51 (60\%)                                                                & 34 (40\%)                                                               & 85                       \\ 
\hline
\textbf{Ambari}                                                                                   & \begin{tabular}[c]{@{}l@{}}System Administrators\\for Hadoop\end{tabular}                            & 34 (47.22\%)                                                             & 38 (52.78\%)                                                            & 72                       \\ 
\hline
\textbf{Hadoop}                                                                                   & \begin{tabular}[c]{@{}l@{}}Distributed Computing\\Environment\end{tabular}                           & 51 (50.50\%)                                                             & 50 (49.50\%)                                                            & 101                      \\ 
\hline
\textbf{Jackrabbit}                                                                               & \begin{tabular}[c]{@{}l@{}}Content Repository \\for Java API\end{tabular}                            & 58 (58.59\%)                                                             & 41 (41.41\%)                                                            & 99                       \\ 
\hline
\textbf{Lucene}                                                                                   & \begin{tabular}[c]{@{}l@{}}Text Search~~Engine\\~Library\end{tabular}                                & 131 (65.83\%)                                                            & 68 (34.17\%)                                                            & 199                      \\ 
\hline
\textbf{Oozie}                                                                                    & \begin{tabular}[c]{@{}l@{}}Executes Hadoop \\Workloads~via \\Web Services\end{tabular}               & 46 (53.49\%)                                                             & 40 (46.51\%)                                                            & 86                       \\ 
\hline
\multicolumn{1}{l}{}                                                                              & \multicolumn{1}{l}{}                                                                                 & \multicolumn{2}{r}{\textbf{Total Commit Instances:~}}                                                                                                       & \multicolumn{1}{c}{\textbf{642}}  \\

\hline
\multicolumn{5}{l}{\begin{tabular}[c]{@{}l@{}}\textit{\textbf{Programming Language:} Java} \end{tabular}}   

\end{tabular}
\end{table}
\begin{table}
\centering
\caption{\textsc{Dataset Summary-II (Automatically Labeled \cite{pornprasit2021jitline})}}
\label{tab:data-summary-ii}
\begin{tabular}{|l|l|c|c|c|c|} 
\hline
\multicolumn{1}{|c|}{\begin{tabular}[c]{@{}c@{}}\textbf{Subject}\\\textbf{ Systems}\end{tabular}} & \multicolumn{1}{c|}{\begin{tabular}[c]{@{}c@{}}\textbf{Application }\\\textbf{Domains}\end{tabular}} & \begin{tabular}[c]{@{}c@{}}\textbf{Prg.}\\\textbf{Lang.}\end{tabular} & \begin{tabular}[c]{@{}c@{}}\textbf{BIC }\\\textbf{ (L = 1)}\end{tabular} & \begin{tabular}[c]{@{}c@{}}\textbf{CC }\\\textbf{ (L = 0)}\end{tabular} & \textbf{BIC+CC}                     \\ 
\hline
\textbf{QT}                                                                                       & \begin{tabular}[c]{@{}l@{}}Cross-platform \\application \\framework\end{tabular}                    & CPP                                                                               & 1,047 (43\%)                                                             & 1,396 (57\%)                                                            & 2,443                               \\ 
\hline
\textbf{OpenStack}                                                                                & \begin{tabular}[c]{@{}l@{}}Open-source \\software \\platform\end{tabular}                            & Python                                                                            & 908 (58\%)                                                               & 663 (42\%)                                                              & 1,571                               \\ 
\hline
\multicolumn{1}{l}{}                                                                              & \multicolumn{1}{l}{}                                                                                 & \multicolumn{3}{r}{\textbf{Total Commit Instances:}}                                                                                                                                                                                  & \multicolumn{1}{c}{\textbf{4,014}} \\

\hline
\multicolumn{6}{l}{\begin{tabular}[c]{@{}l@{}}\textit{\textbf{BIC:} Bug Inducing Commits}\\\textit{\textbf{CC:} Clean Commits (Commits that fix a bug and no more bug is reported)}\\\textit{\textbf{L:} Data Label}\end{tabular}}   

\end{tabular}
\end{table}

Our key contribution to this investigation is to propose source code syntax pattern-based features to detect bug-inducing commits (BICs) using different machine learning (ML) models. We compare the performance of detecting BICs using the combinations of commonly used feature values obtained from similar studies \cite{2019:Borg:SZZUnleashed, CommitGuruRosen, Rel:Fukushima:2014:ESJ:2597073.2597075} and our proposed feature values. We want to verify whether our proposed features extracted from source code syntax patterns of software projects improve the performance of ML models in detecting BICs. There are a lot of different studies \cite{Kim:Classify:Clean:Buggy, Rel:Kim:2007:PFC:1248820.1248881, Rel:RePEc:ids:ijrsaf:v:7:y:2013:i:1:p:17-31, Rel:2013:Shivaji:Reducing:Feature:Bug:Prediction,  Rel:Kim:2006:AIB:1169218.1169308, Sliwerski-2005-induce-fixes-journal, Rel:Wen:2016:LLB:2970276.2970359} in the literature using similar types of feature values containing different experimental setups (i.e., machine learning, deep learning \cite{Rel:2015:Yang:Deep:JIT:Defect:Prediction, DeepJITHowFar}, single project, cross-project \cite{Rel:Fukushima:2014:ESJ:2597073.2597075, ICSE2020-TransferLearningCrossJIT, ICSE2020-CrossJIT, AEHasan-JIT-Patch}, etc.). Zeng et al. \citet{DeepJITHowFar} replicated the results of two popular Just In Time defect prediction models CC2Vec \cite{ICSE2020-CC2VEC} and DeepJIT \cite{DeepJIT:MSR:2019}. They reported the difference in obtained results between the original datasets and the experimental projects used in their study. They show that the selection of datasets and feature values impact the CC2Vec and DeepJIT defect prediction models. As in this investigation, we propose new types of features extracted from the source code syntax pattern of software projects to be used in ML models. We also aim to use six manually and two automatically labeled datasets of the buggy and non-buggy commit to performing the investigation. Instead of directly using the findings of any other study, we reproduce all the results in our experimental setup using commonly used feature values (which we give the name GitHub Statistics or GS) and compare them with the results of our proposed features. Therefore, we took our results using commonly used features in recent studies as the baseline result and compared them with the results using different combinations of our proposed features. As we wanted to see an improvement in detecting buggy and non-buggy commits using different feature combinations, we generated all the results in our implementation setup for a fair comparison among them. Our investigation shows that using the features extracted from the source code syntax patterns improves the performance of detecting BICs. We believe that any other experimental setup should provide a similar result. In the future, we plan to replicate recent studies and compare the improvement of buggy commit detection using our proposed features extracted from the source code syntax pattern.

Our baseline is the result we obtain using conventional feature values (GS-ALL, GS) to identify bug-inducing commits similar to the existing related works \cite{2019:Borg:SZZUnleashed, CommitGuruRosen, Rel:Fukushima:2014:ESJ:2597073.2597075}. We then compare the results using developers’ coding syntax pattern-based features (TS, TP) and their various possible combinations in detecting BIC from eight subject systems. 
The token patterns and sequences generated in Figure \ref{fig:SourceFeatureExtractionSrcML} demonstrate their effects on the control flow of the source code. A simple token pattern represents a simple coding structure, and a simple code is more likely to be easier for software developers to edit and less likely to be bug-inducing. We investigate the encoding of these patterns as feature values to detect bug-inducing  commits in this study. We also investigate the set of best features for each of the individual subject systems.         

We applied Random Forest (RF) Classifier to the datasets of the six software repositories shown in Table \ref{tab:data-summary-i} and five different ML-based classification models to the datasets of the two software repositories shown in Table \ref{tab:data-summary-ii} to detect bug-inducing commits and evaluated our results to answer the following research questions.

\vspace{0.25cm}
\noindent
\textbf{RQ1:} How can we determine whether developers’ coding syntax patterns could be responsible for inducing bugs in a software system?
\begin{itemize}
\item We encoded developers’ coding syntax patterns as feature values and named them Token Sequence (TS) and Token Pattern (TP). Results of our investigation show that the inclusion of TS and TP as feature values provides better bug-inducing commit (BIC) detection performance than the conventional features (GS-ALL and GS). Therefore, we can conclude that, as TS and TP features contribute to detecting BIC using ML models, they might also be responsible for inducing bugs in software projects. Some repeating patterns might induce bugs in software systems, and ML models are being trained to identify those patterns utilizing the proposed TS and TP features. 
\end{itemize}

\vspace{0.25cm}
\noindent
\textbf{RQ2:} Do the features extracted from developers' coding syntax patterns provide significantly better performance compared to the other feature values?
\begin{itemize}
\item We perform the Wilcoxon Signed Rank test \cite{wilcoxon-signed-rank-test, wilcoxon-signed-rank-test-rosner} with the results we obtained in manually labeled six subject systems. Our investigation shows the improvements in precision, recall, and f1~scores are statistically significant.  
\end{itemize}

\vspace{0.25cm}
\noindent
\textbf{RQ3:} How generalized are the extracted features from one software system to the others?
\begin{itemize}
 \item Our findings show the number of features is different in the collection of best features in different subject systems if the dataset size is not large enough (our manually labeled six subject system). Based on the findings of this study, we can say that as different developers maintain different software systems, their coding syntax styles, working patterns, and codebase size are also different. Therefore, we require a separate list of features to detect future occurrences of buggy commits from those systems. With large datasets (automatically labeled two subject systems), we get improved BIC detection results using all the feature values without prioritization in five different ML detection models. 
\end{itemize}

\vspace{0.25cm}
\noindent
\textbf{RQ4:} Do the features extracted from  developers’ coding syntax patterns enhance the explainability of BIC detection from software systems?
\begin{itemize}
\item We utilized PyExplainer \cite{CitePyExplainer}, a machine learning prediction explaining tool. The use of this tool provided us with some examples in Table 7, which clearly demonstrates that the use of source code syntactic pattern-based features provides better reasoning compared to the code-churn-based features about the bug proneness of a software commit. 
\end{itemize}

The knowledge of this study will help avoid future bug introduction in the codebase, which will bring effective financial and resource utilization to the software development community. The source code of commit patches and their equivalent XML representation from all the eight subject systems we use in this study are publicly available\footnote{\label{bicDetectionSF}https://github.com/mnadims/bicDetectionSF} for readers to investigate and facilitate any replication study.  

We organized this paper into the following sections. A full overview of this study in different steps is in section \ref{study-overview}; the results of this study are described and analyzed in section \ref{result-discussion}. Threats to validity are described in section \ref{threats-validity}, and section \ref{related-work} performs the literature review of existing similar studies. Finally, we conclude this paper in section \ref{conclusion-future-work}, mentioning the future directions of this study.

\section{Study Overview}
\label{study-overview}
Most of the earlier studies to detect BICs either used features only from GitHub statistical data \cite{2019:Borg:SZZUnleashed, CommitGuruRosen} or combined GitHub statistical features with natural language-based text processing of source code \cite{ReducingFeatures-BIC} in Machine Learning Models. A systematic literature review \cite{fault-prediction-study-survey} reports that different studies use features extracted from various sources, such as previous historical data from GitHub or submitted bug reports, source code metrics, or any combination of those features. For example, some studies \cite{ DeepJIT:MSR:2019, AEHasan-JIT-Patch} use historical patches to identify defects from software systems, and they call it JIT (Just in Time defect prediction). \citet{fault-prediction-study-survey} also find that only source code metric works very poorly if no other kinds of features (such as GitHub Statistics) are combined with that. The decision to take which values to be used as features depends on some underlying assumptions. One such commonly used feature is a Line of codes (LOC) in almost all ML-based software quality assurance or defect prediction studies \cite{fault-prediction-study-survey}. A bigger software system (more LOC) is more likely to be buggy than a smaller one is the underlying assumption of using LOC as a feature value. Some other standard features (such as Lines of Code Added/ Deleted/ Updated, Number of Directories Modified, and Experience of a Developer)  used in these studies are also based on similar assumptions. These features can provide an assumption or likelihood of a commit being buggy and will always remain far behind to give a clear idea about which source code portion has introduced the bug or how we will find or fix it. We should focus on maintenance activities where source code is added or updated to identify specific reasons for bug introduction within a particular software system. 

\begin{figure}
\centering
\includegraphics[width=\textwidth] {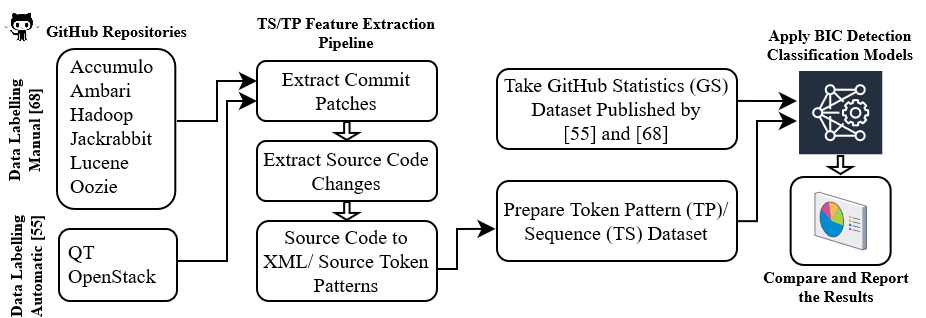}
\caption{Overall Steps of this Study}
\label{fig:BICOverall}
\end{figure}

We perform this study to verify whether the developers' coding syntax patterns can induce a bug in a software system by utilizing developers' coding syntax pattern-based features (TS and TP) in ML-based detection models. We demonstrate the overall study in Figure \ref{fig:BICOverall}, which includes the following key steps.

\subsection{\textbf{Datasets Labelling (Buggy and Non-buggy Commits)}}
\label{labeling-dataset}
To evaluate the efficiency of any techniques in detecting Bug Inducing Commits (BICs) or Just in Time (JIT) defect prediction, most of the related studies \cite{2019:Borg:SZZUnleashed, CommitGuruRosen} used datasets labeled using the SZZ algorithm \cite{Sliwerski-2005-induce-fixes-journal}. \citet{2019:FSE:Wen:EEC:3338906.3338962} reported that about 63.7\% of bug-inducing  commits (BIC) identified by the SZZ Algorithm are not real. There are several updated versions of the SZZ algorithm to deal with its incorrect labeling  \cite{Framework:SZZ, Text:Based:Bug:Inducing, Rel:Kim:2006:AIB:1169218.1169308}. We do not use a dataset labeled using the SZZ algorithm in this study because of its imprecision. As our primary goal is to verify the effectiveness of features extracted from developers’ coding syntax patterns, we utilize labeled datasets published by recent similar studies as our baseline and added our TS and TP-based features to investigate the improvements in detecting bug inducing commits (BICs). Data sets selected in this investigation fall into the following categories:  

\subsubsection{\textbf{Manually labeled Datasets (Table \ref{tab:data-summary-i})}}
Six of the eight datasets used in this investigation are labeled manually by \citet{2019:FSE:Wen:EEC:3338906.3338962}, which is publicly available \footnote{https://github.com/justinwm/InduceBenchmark} to access. The datasets of bug-fixing commits and the associated bug-inducing commits for seven different projects were used in their paper. We utilized six available project data files as they contain enough data instances to apply Machine Learning based classification models such as Random Forest Classifier. We could not utilize one project data file (./Defects4J.csv) as it contains only the bug-inducing commits and associated Bug IDs from five different projects. In addition, it includes 92 data instances from five various projects, making insufficient data instances to apply ML-based classification models on the dataset of each of these projects.

\subsubsection{\textbf{Automatically Labeled Datasets (Table \ref{tab:data-summary-ii})}}
Our investigation also utilizes two popular datasets, QT\footnote{https://github.com/qt/} and OpenStack\footnote{https://github.com/openstack/}, used in similar studies \cite{DeepJIT:MSR:2019, pornprasit2021jitline}, which are labeled automatically and publicly available to access. We first download their data files from their data repository and then extract the commit instances which have a labeled number of bugs. We find the labeled bug presence falls into two categories, those which have a bug count of 0, and those which have 1 or more bugs. To make it simpler for buggy commit detection by binary classification and compatible with our other six subject systems, we labeled the commit instances from these two subject systems which have 1 or more bugs as buggy commit instances and labeled them L=1. The commits which do not have any bugs are labeled as L=0. OpenStack and QT contain 754 and 89 repositories, respectively, as their sub-modules. We find the source code patches of the labeled commits from QT in a sub-module named 'qtbase'. The labeled commits of OpenStack are from different sub-modules. To extract the source code patches, we first search each commit in all the 754 sub-modules of OpenStack and then store the source code of that commit patch for further processing. 

The datasets from the manually labeled six projects are shown in Table \ref{tab:data-summary-i}, and two automatically labeled projects are in Table \ref{tab:data-summary-ii}.  We investigate the BIC detection performance of the ML model using our introduced features on these six project datasets. There are two categories of commits in the dataset; one is identified as bug-inducing  commits, and the others are identified as bug-fixing commits by manual investigation. We labeled those bug-inducing  commits as L=1. As training and testing of an ML-based detection model require the target class and its different classes to provide as training data samples, we labeled the bug fixing commits, which are not inducing new bugs in the system, as clean commits (L=0). The data samples used to train and test the ML model those commits are mainly removing the problem/ issues inserted by the commits, which are labeled as bug-inducing  (L=1). These commits can provide more distinguishable features to ML-based detection models to identify two categories of commits. 

\subsection{\textbf{Identifying Developers' Coding Syntax Patterns and Sequences}}
\label{detailTS-TP}
Developers' coding syntax pattern represents the behavior of a software developer, which affects the overall structure of a source code statement. We identify thousands of different commit patterns from source code patches of the eight subject systems of this investigation. About 18k of these patches whose token pattern length is within 100 characters are publicly available to view by the readers at our GitHub repository\textsuperscript{\ref{bicDetectionSF}} for an easier understanding of those patterns. We also extract a sequence of these patterns for preparing feature values, which could be used in our ML-based classification model. As tokens of a programming language are the basic building blocks of each feature type extracted from the XML representation of source codes, we name the categories of features as Token Sequence (TS) and Token Pattern (TP). Commits encoded by these TS and TP represent the syntax style of coding by different developers. 

\begin{figure}
\includegraphics[width=\textwidth] {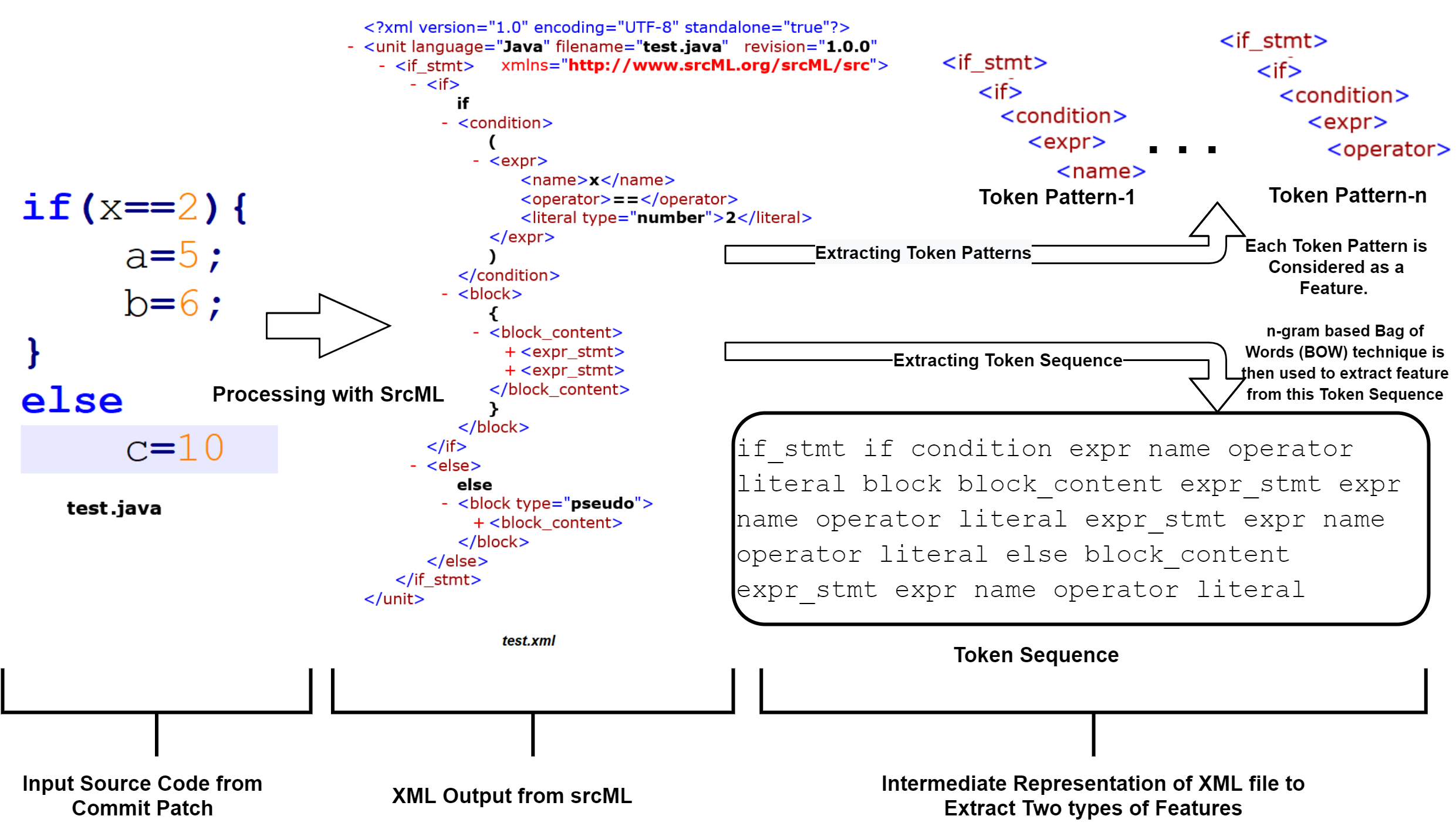}
\caption{Demonstration of Feature Extraction from the Source Code of Commit Patch}
\label{fig:SourceFeatureExtractionSrcML}
\end{figure}

We show an overall demonstration of extracting features (TS and TP) using a straightforward piece of program in Java language in Figure \ref{fig:SourceFeatureExtractionSrcML}. Few additional examples of more complex program constructions, such as adding AND, OR in if-statement, nested if statement, and switch-case statement, are also available to view in four different other files ($./simpleDemonstrations$) at our GitHub repository\textsuperscript{\ref{bicDetectionSF}}. Our public repository also contains all the project files ($./projectFiles$), including the source file of commit patches and their equivalent XML representations for further investigation and reproduction of this study.

We first downloaded all the commit patches using the commit id in our labeled dataset and the git command for corresponding GitHub repositories. Then we retrieved the source code segments from those patch files. After that, we utilized a tool (srcML\footnote{https://www.srcml.org/}), which converts source code into an XML format representing its coding structure in a normalized form. Suppose we have a source code of a commit-patch in a file named test.java, as shown in Figure \ref{fig:SourceFeatureExtractionSrcML}. Processing the file test.java with the srcML tool will generate an XML file test.xml. The XML representation of the source file provides a hierarchy of the source code statement structure and token organization. This example shows that the file test.java contains two main blocks, one for $if$ and the corresponding $else$ for the $if$. In the XML representation of the source file, we can see those two blocks keeping the identifier hierarchy and sequence similarity. Processing the XML file, we can extract the normalized names of different tokens in the source code and the hierarchy of those tokens. Examples of the extracted token sequence (TS) and token pattern (TP) are shown in Figure \ref{fig:SourceFeatureExtractionSrcML}.  We have given each token pattern an id such as $pattern\_id\_1,~pattern\_id\_2, ~$.......$ ~pattern\_id\_n$ where n is the total number of patterns extracted from the source file. Here, $pattern\_id\_1$ means the extracted pattern of normalized tokens, such as \textbf{if\_stmt-if-condition-expr-name} is one of the patterns shown in Figure \ref{fig:SourceFeatureExtractionSrcML}. There could be thousands of such patterns extracted from a software system that will reflect how the developer of that software system performs coding. To prepare the dataset for a commit, we will consider how many such patterns are in the source code of that commit patch. Similarly, in the same figure, we can also see an example of the extracted token sequence. We processed that token sequence with the n-gram \cite{cavnar1994n-gram} technique, which is mainly used in extracting features from natural language text. We utilized n-gram (with n=1 to n=5) in our extracted token sequence to produce Bag of Word (BOW) \cite{bagofwords, DistributedRepresentationDocument, FuzzyBOW}, which is then used as another source of features, and we named these features as Token Sequence or TS. We repeated this process for the source code of every commit patch in all the subject systems to extract the features from TS and TP. 

\subsection{\textbf{Encoding Developers' Coding Syntax Patterns to be used in the ML-based Detection Model}}
\label{encoding-ts-tp}
Applying ML-based models depends on the availability of labeled datasets (sample commits from each of the labeled classes) and extracting relevant feature values from the codebase history to train and test the model. After identifying the TS and TP patterns from the source code of the commit patch, it is important to encode them as feature values, which could be used in ML-based detection models. In this study, we used simple encoding of TS and TP by considering their number of occurrences in the source code of the commit patch. For example, in Figure \ref{fig:SourceFeatureExtractionSrcML}, we demonstrated how the TS and TP patterns are extracted from the Java source code. Each code fragment may have several TS and TP.  We extract all the TS and TP from all the commit patches of our labeled datasets. After that, we selected unique TS and TP to be used as feature identifiers. We will use the count of those feature identifiers in each commit to represent that commit while using ML-based detection models. A sample of TP is extracted for each commit, and their encoding mechanism is demonstrated in Table \ref{tab:sample-tp} and Table \ref{tab:sample-encoding-tp}. In Table \ref{tab:sample-encoding-tp}, we show the commit id, its extracted TP, and the associated label for each of the commits. Here, L=0 indicates a clean commit, and L=1 indicates a BIC. We then took unique TP from all the TPs to encode them and took the number of occurrences of each TP to represent each of the commits in Table \ref{tab:sample-encoding-tp}.  

\begin{table}[ht]
\centering
\caption{\textsc{Sample Token Pattern (TP) in Each Commit}}
\label{tab:sample-tp}
\begin{tabular}{|c|l|c|} 
\hline
\textbf{Commit ID} & \multicolumn{1}{c|}{\textbf{Token Pattern (TP)}} & \textbf{Label (L)}  \\ 
\hline
1                  & TP1, TP2, TP3, TP3, TP4, TP4                     & 0                   \\ 
\hline
2                  & TP2, TP4, TP5                                    & 1                   \\ 
\hline
3                  & TP1, TP3, TP6                                    & 0                   \\ 
\hline
4                  & TP4, TP4, TP4                                    & 0                   \\ 
\hline
5                  & TP4, TP6, TP7, TP7                               & 1                   \\
\hline
\end{tabular}
\end{table}

\begin{table}[ht]
\centering
\caption{\textsc{Sample Encoding of Token Patterns (TP) of Table \ref{tab:sample-tp}}}
\small\addtolength{\tabcolsep}{-2pt}
\label{tab:sample-encoding-tp}
\begin{tabular}{|c|c|c|c|c|c|c|c|c|} 
\hline
\begin{tabular}[c]{@{}c@{}}\textbf{Commit }\\\textbf{ID}\end{tabular} & \textbf{TP1} & \textbf{TP2} & \textbf{TP3} & \textbf{TP4} & \textbf{TP5} & \textbf{TP6} & \textbf{TP7} & \begin{tabular}[c]{@{}c@{}}\textbf{Label }\\\textbf{(L)}\end{tabular}  \\ 
\hline
1                                                                     & 1            & 1            & 2            & 2            & 0            & 0            & 0            & 0                                                                      \\ 
\hline
2                                                                     & 0            & 1            & 0            & 1            & 1            & 0            & 0            & 1                                                                      \\ 
\hline
3                                                                     & 1            & 0            & 1            & 0            & 0            & 1            & 0            & 0                                                                      \\ 
\hline
4                                                                     & 0            & 0            & 0            & 3            & 0            & 0            & 0            & 0                                                                      \\ 
\hline
5                                                                     & 0            & 0            & 0            & 1            & 0            & 1            & 2            & 1                                                                      \\
\hline
\end{tabular}
\end{table}

 
\subsection{\textbf{Selecting The Set of Best Features for each of the Subject Systems}}
\label{selecting-feature-ranking}
\begin{figure}
\centering
\includegraphics[width=0.75\textwidth] {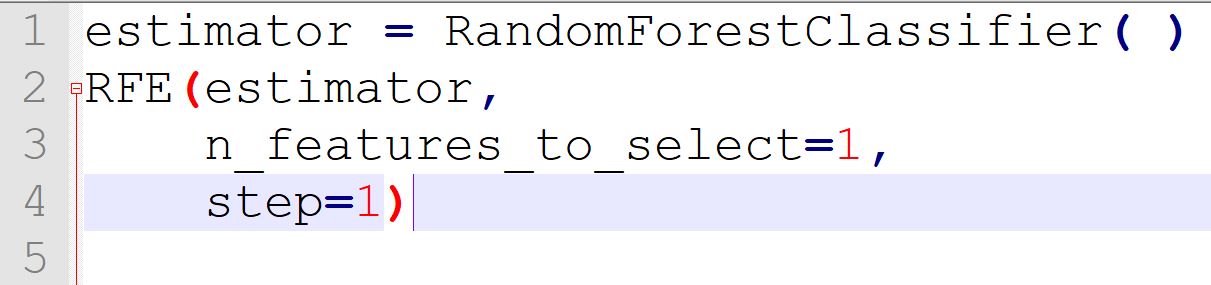}
\caption{Recursive Feature Elimination}
\label{lst:RFE}
\end{figure}

Machine learning algorithms provide its result based on the feature values given for each data instance. The features we are considering for training and testing a machine learning model play a vital role in the performance of that model. A weak or irrelevant feature list can degrade the overall accuracy of classification. We wanted to evaluate whether the best features for a subject system are also better for the other subject systems. We applied Recursive Feature Elimination (RFE) of SciKit Learn \cite{scikit-learn} python library to rank all the features in the dataset of a subject system. RFE is a recursive approach to finding a list of the best features for a dataset with its classification labels. A user first needs to specify the number of best features expected to keep and the number of features to eliminate in each step. RFE then eliminates the weakest feature (or features) in each recursive step until the expected number of features is obtained. The parameter settings we have used for RFE are shown in Figure \ref{lst:RFE}, where we used the Random Forest Classifier to estimate the importance of features and eliminate one feature in each step.


As we set $n\_features\_to\_select=1$, it will rank all the features from rank 1 to n (n=total number of features). Then return the top feature as the best feature for the dataset classification. We performed this for each feature combination, such as GS, TS, TP, GS+TS, etc. As GS has 12 features, this step will rank those features from rank-1 to rank-12 based on their importance in classifying the dataset. Similarly, GS+TS features have $12+10514=10526$ features, and they will get from rank-1 to rank-10526. 

After obtaining the ranking of each of the features from the RFE algorithm, we applied the Random Forest classifier to select the set of best features for each of the subject systems. We first determine the Precision, Recall, and F1 Score of BIC detection using only the top (rank-1) feature in a subject system. Then we added one feature in each step from the following position in the rank list and determined the classification performance. We discarded the added feature if it did not improve the performance obtained in the previous step. Therefore, after completing the iteration from feature 1 to the total number of features, we will have the best features for that subject system. We repeated this approach for each of the feature combinations and each of the six manually labeled subject systems. 

\subsection{\textbf{Applying ML-based BIC Detection Models}}
\label{ml-model}
Our investigation utilizes five machine learning classification models to detect bug inducing commits from the two (OpenStack and QT) of the eight subject systems. We use OpenStack and QT datasets for applying multiple ML models as these datasets are most used in similar studies and contain a large number of data samples compared to the other manually labeled six subject systems. We use only the Random Forest (RF) classification model on the manually labeled datasets as they contain a smaller number of data instances. We find the features extracted from source code statements improve the prediction of defective software commits. All the classification models except the DBN are available in the Python SciKit Learn \cite{scikit-learn} library. We utilized the DBN classification model made available by \citet{DBNAlbert}. 
\begin{enumerate}[i.]
  \item \textbf{Random Forest (RF)} Classifier implements ensemble learning methods where it builds multiple decision trees and merges them to get a better and more stable prediction. At the root of each tree, it first divides the data in such a way that the differences in each part of the data become as lower as possible, then it again divides the branches of the tree to get more specialization on the data. We can control the depth of such a tree by the parameter variable \textbf{max\_depth}. The algorithm repeats this process with some different random selections and creates a set of random decision trees. During prediction, it predicts the case using each of the decision trees in the forest, then finalizes the decision with the majority decision of the trees. 
  \item \textbf{K-Nearest Neighbors (KNN)} Classifier is one of the most commonly used ML algorithms known for its simplicity and effectiveness \cite{KNNAlgo}. It is a supervised learning method that uses a labeled training dataset to categorize data points for predicting the category of the test dataset. 
  \item \textbf{Gradient Boosting Classifier (GBC)} is also an ensemble learning method, where the model learns by optimizing the loss function. It is used for both classification and regression. We use the default configuration of the Python SciKit Learn \cite{scikit-learn} library for utilizing this algorithm to detect bug inducing commits. 
  \item \textbf{Perceptron (PCT)} is the simplest type of neural network model used for binary classification. It takes a row of input data to predict its output class. It utilizes the weighted sum of the input and a bias value. During the training process of PCT, it updates its weights to minimize the error of classification in training datasets. 
  \item \textbf{Supervised Deep Belief Network (DBN)} Classification is the utilization of an advanced deep learning algorithm \cite{Rel:2015:Yang:Deep:JIT:Defect:Prediction, DBN1, DBN2, DBN3} for detecting more meaningful features to increase the classification performance of a labeled dataset. It contains two phases, i) feature selection, and ii) machine learning based classification. In the feature selection phase, it utilizes provided features to extract more meaningful feature values and then applies classification based on the extracted features. 
 \end{enumerate}

We have applied machine learning models using time-sensitive detection techniques, which means training data must be from past timestamps than the testing data.  \citet{Tan:2015:ODP:ICSE} show that cross-validation can provide false higher precision in predicting a data instance that is sensitive to timestamp, and a situation may happen then train the ML model with the data from the future and tests it using the data from the past. We first sorted the dataset in ascending order based on the commit date and took the first 70\% of the data for training the model and the remaining 30\% for testing. All the commit instances in the training dataset are from the past compared to the testing data instances. 

\subsection{\textbf{Feature Combinations}}
\label{feature-combinations}
As we extracted Token Pattern (TP) features utilizing the SrcML tool, which does not support Python for converting the source fragments into an XML representation. Therefore, we extracted only token sequence (TS) features from OpenStack as it is written using Python programming language. We extract both the TS and TP features from all the other seven subject systems listed in Tables \ref{tab:data-summary-i} and \ref{tab:data-summary-ii}. We investigated the ML-based classification model's performance by taking different combinations of GS, TS, and TP features. In this investigation, our baseline feature is GS, and we are proposing TS and TP features to improve BIC detection performance. Suppose we would like to combine GS and TS features for the subject system Accumulo where the number of GS and TS features is 12 and 10,514. Thus, the number of combined total features (GS+TS) in Accumulo will be 10,526. We will rank those features based on their importance, where the most important feature will get rank-1, and the least important feature will get rank-10526. Similarly, if we would like to investigate GS+TS+TP for Accumulo, we will have 12+10,514+741= 11,267 features and rank them from 1 to 11,267th based on their importance in the dataset classification of Accumulo.  As we have three basic feature types, we got seven feature combinations (GS, TS, TP, GS+TS, GS+TP, TS+TP, GS+TS+TP). We wanted to investigate whether more than one type of feature improved the classification performance by combining different feature types. Each of the ranked seven feature combinations is prioritized using the technique described in Section \ref{selecting-feature-ranking}.  


\begin{table}
\centering
\caption{\textsc{Performance of Single Type Features}}
\label{tab:single-features-final-result}
\begin{tabular}{|l|l|c|c|c|c|} 
\hline
\multicolumn{1}{|c|}{\begin{tabular}[c]{@{}c@{}}\textbf{Feature}\\\textbf{ Type}\end{tabular}}                                                              & \multicolumn{1}{c|}{\begin{tabular}[c]{@{}c@{}}\textbf{Subject}\\\textbf{ System}\end{tabular}} & \textbf{Precision} & \textbf{Recall} & \textbf{F1 Score} & \textbf{AUC}  \\ 
\hline
\multirow{6}{*}{\begin{tabular}[c]{@{}l@{}}\textbf{GitHub }\\\textbf{Statistics }\\\textbf{All Features}\\\textbf{(GS-ALL)}\end{tabular}}                & Accumulo                                                                                        & 0.57               & 0.73            & 0.64              & 0.76          \\ 
\cline{2-6}
                                                                                                                                                            & Ambari                                                                                          & 0.67               & 0.80            & 0.73              & 0.76          \\ 
\cline{2-6}
                                                                                                                                                            & Hadoop                                                                                          & 0.40               & 0.91            & 0.56              & 0.75          \\ 
\cline{2-6}
                                                                                                                                                            & Jackrabbit                                                                                      & 0.50               & 0.69            & 0.58              & 0.55          \\ 
\cline{2-6}
                                                                                                                                                            & Lucene                                                                                          & 0.63               & 0.79            & 0.70              & 0.78          \\ 
\cline{2-6}
                                                                                                                                                            & Oozie                                                                                           & 0.50               & 0.67            & 0.57              & 0.64          \\ 
\hline
\multirow{6}{*}{\begin{tabular}[c]{@{}l@{}}\textbf{GitHub }\\\textbf{Statistics}\\\textbf{Prioritized }\\\textbf{Features}\\\textbf{(GS)}\end{tabular}} & Accumulo                                                                                        & 0.69               & 0.82            & 0.75              & 0.73          \\ 
\cline{2-6}
                                                                                                                                                            & Ambari                                                                                          & 0.89               & 0.80            & 0.84              & 0.92          \\ 
\cline{2-6}
                                                                                                                                                            & Hadoop                                                                                          & 0.53               & 0.91            & 0.67              & 0.78          \\ 
\cline{2-6}
                                                                                                                                                            & Jackrabbit                                                                                      & 0.53               & 0.77            & 0.62              & 0.55          \\ 
\cline{2-6}
                                                                                                                                                            & Lucene                                                                                          & 0.73               & 0.82            & 0.77              & 0.83          \\ 
\cline{2-6}
                                                                                                                                                            & Oozie                                                                                           & 0.56               & 0.83            & 0.67              & 0.65          \\ 
\hline
\multirow{6}{*}{\begin{tabular}[c]{@{}l@{}}\textbf{Token }\\\textbf{Sequence }\\\textbf{(TS)}\end{tabular}}                                               & Accumulo                                                                                        & 0.69               & 1.00            & 0.81              & 0.81          \\ 
\cline{2-6}
                                                                                                                                                            & Ambari                                                                                          & 0.75               & 0.60            & 0.67              & 0.77          \\ 
\cline{2-6}
                                                                                                                                                            & Hadoop                                                                                          & 0.85               & 1.00            & 0.92              & 0.96          \\ 
\cline{2-6}
                                                                                                                                                            & Jackrabbit                                                                                      & 0.65               & 1.00            & 0.79              & 0.82          \\ 
\cline{2-6}
                                                                                                                                                            & Lucene                                                                                          & 0.82               & 1.00            & 0.90              & 0.92          \\ 
\cline{2-6}
                                                                                                                                                            & Oozie                                                                                           & 0.86               & 1.00            & 0.92              & 0.91          \\ 
\hline
\multirow{6}{*}{\begin{tabular}[c]{@{}l@{}}\textbf{Token }\\\textbf{Pattern }\\\textbf{(TP)}\end{tabular}}                                                & Accumulo                                                                                        & 0.58               & 1.00            & 0.73              & 0.88          \\ 
\cline{2-6}
                                                                                                                                                            & Ambari                                                                                          & 0.90               & 0.90            & 0.90              & 0.91          \\ 
\cline{2-6}
                                                                                                                                                            & Hadoop                                                                                          & 0.64               & 0.82            & 0.72              & 0.67          \\ 
\cline{2-6}
                                                                                                                                                            & Jackrabbit                                                                                      & 0.65               & 1.00            & 0.79              & 0.66          \\ 
\cline{2-6}
                                                                                                                                                            & Lucene                                                                                          & 0.67               & 0.94            & 0.78              & 0.72          \\ 
\cline{2-6}
                                                                                                                                                            & Oozie                                                                                           & 0.86               & 1.00            & 0.92              & 0.91          \\
\hline
\end{tabular}
\end{table}

\begin{table}
\centering
\caption{\textsc{Performance of Multiple Type Feature Combinations}}
\label{tab:combined-features-final-result}
\begin{tabular}{|l|l|c|c|c|c|} 
\hline
\multicolumn{1}{|c|}{\begin{tabular}[c]{@{}c@{}}\textbf{Combined}\\\textbf{ Features}\end{tabular}} & \multicolumn{1}{c|}{\begin{tabular}[c]{@{}c@{}}\textbf{Subject}\\\textbf{ System}\end{tabular}} & \textbf{Precision} & \textbf{Recall} & \textbf{F1 Score} & \textbf{AUC}  \\ 
\hline
\multirow{6}{*}{\textbf{GS+TS}}                                                                     & Accumulo                                                                                        & 0.77               & 0.91            & 0.83              & 0.91          \\ 
\cline{2-6}
                                                                                                    & Ambari                                                                                          & 0.75               & 0.60            & 0.67              & 0.72          \\ 
\cline{2-6}
                                                                                                    & Hadoop                                                                                          & 0.56               & 0.91            & 0.69              & 0.77          \\ 
\cline{2-6}
                                                                                                    & Jackrabbit                                                                                      & 0.61               & 0.85            & 0.71              & 0.62          \\ 
\cline{2-6}
                                                                                                    & Lucene                                                                                          & 0.94               & 0.97            & 0.96              & 0.97          \\ 
\cline{2-6}
                                                                                                    & Oozie                                                                                           & 0.71               & 1.00            & 0.83              & 0.74          \\ 
\hline
\multirow{6}{*}{\textbf{GS+TP}}                                                                     & Accumulo                                                                                        & 0.71               & 0.91            & 0.80              & 0.82          \\ 
\cline{2-6}
                                                                                                    & Ambari                                                                                          & 0.82               & 0.90            & 0.86              & 0.76          \\ 
\cline{2-6}
                                                                                                    & Hadoop                                                                                          & 0.62               & 0.91            & 0.74              & 0.76          \\ 
\cline{2-6}
                                                                                                    & Jackrabbit                                                                                      & 0.71               & 0.92            & 0.80              & 0.78          \\ 
\cline{2-6}
                                                                                                    & Lucene                                                                                          & 0.79               & 0.91            & 0.85              & 0.83          \\ 
\cline{2-6}
                                                                                                    & Oozie                                                                                           & 0.73               & 0.92            & 0.81              & 0.74          \\ 
\hline
\multirow{6}{*}{\textbf{TS+TP}}                                                                     & Accumulo                                                                                        & 0.73               & 1.00            & 0.85              & 0.81          \\ 
\cline{2-6}
                                                                                                    & Ambari                                                                                          & 0.75               & 0.60            & 0.67              & 0.77          \\ 
\cline{2-6}
                                                                                                    & Hadoop                                                                                          & 0.67               & 0.91            & 0.77              & 0.77          \\ 
\cline{2-6}
                                                                                                    & Jackrabbit                                                                                      & 0.68               & 1.00            & 0.81              & 0.78          \\ 
\cline{2-6}
                                                                                                    & Lucene                                                                                          & 0.82               & 1.00            & 0.90              & 0.90          \\ 
\cline{2-6}
                                                                                                    & Oozie                                                                                           & 0.67               & 1.00            & 0.80              & 0.78          \\ 
\hline
\multirow{6}{*}{\textbf{GS+TS+TP}}                                                                  & Accumulo                                                                                        & 0.77               & 0.91            & 0.83              & 0.93          \\ 
\cline{2-6}
                                                                                                    & Ambari                                                                                          & 0.80               & 0.80            & 0.80              & 0.75          \\ 
\cline{2-6}
                                                                                                    & Hadoop                                                                                          & 0.53               & 0.91            & 0.67              & 0.66          \\ 
\cline{2-6}
                                                                                                    & Jackrabbit                                                                                      & 0.69               & 0.85            & 0.76              & 0.68          \\ 
\cline{2-6}
                                                                                                    & Lucene                                                                                          & 0.84               & 0.94            & 0.89              & 0.92          \\ 
\cline{2-6}
                                                                                                    & Oozie                                                                                           & 0.86               & 1.00            & 0.92              & 0.90          \\
\hline
\end{tabular}
\end{table}


\section{Results and Discussion}
\label{result-discussion}
Our investigation contains two categories of datasets, i) Manually labeled data instances in Table \ref{tab:data-summary-i} and ii) Automatically labeled data instances in Table \ref{tab:data-summary-ii}. Identification of buggy and non-buggy commit instances is made in both datasets by earlier published studies \cite{pornprasit2021jitline, 2019:FSE:Wen:EEC:3338906.3338962}. We evaluate the performance of detecting bug inducing commits (BICs) using our proposed features (TS and TP) in all eight software projects and compare the result with conventional (GS) features used in earlier studies. We also utilize combining the features to detect BIC from different subject systems. Our results are in Table \ref{tab:single-features-final-result}, Table \ref{tab:combined-features-final-result}, Figure \ref{fig:bicFEngPrecision}, Figure \ref{fig:bicFEngRecall}, and Figure \ref{fig:bicFEngF1Score}. We discuss our results in the following sub-sections. 

\subsection{Evaluation Metric}
We first calculate some metric values to compare the performance of different feature combinations proposed in this study. Calculating these metric values depends on the counts of true positives (TP), true negatives (TN), false positives (FP), and false negatives (FN). These counts indicate how well the machine learning (ML) model detects bug-inducing commits (BICs). For example,  TP and TN are the counts of correctly identified BIC and non-BIC. Similarly, FP and FN indicate the incorrectly detected BIC and non-BIC using the ML model. Once we have these counts for all the feature combinations from the dataset of all the subject systems, we use the following formula to calculate Precision, Recall, and F1~Score.

When an ML model has a lower count of FP, it provides a higher value for precision, indicating the higher reliability of the model. Similarly, a model with lower FN provides a higher recall value indicating higher accuracy of the detected results. However, an attempt to increase the precision value typically decreases the precision and vice versa \cite{precisionRecall}. A BIC detection model that provides a higher precision with lower recall or higher recall with lower precision is not acceptable in a real-life practical software development industry. Therefore, we calculate the F1~Score by taking the harmonic mean of precision and recall to evaluate the results of this study. A higher F1~Score in a comparison scenario indicates a better result, considering the precision and recall values. 
\vspace{1cm}

\[ Precision = \frac{TP}{TP + FP}\]
\[ Recall = \frac{TP}{TP + FN}\]
\[ F1~Score = \frac{2 \times Precision \times Recall}{Precision + Recall}\]


We also calculate the AUC (Area Under the ROC Curve) \cite{ROC:AUC} score, which provides an aggregate performance measure across all possible classification thresholds. The value of the AUC score is between 0 to 1, where 1.0 represents a model 
whose all detections (100\%) are correct; one whose all of the detection are incorrect has an AUC of 0.0.

We use Scikit-learn \cite{scikit-learn} library available in Python programming language to calculate Precision, Recall, F1 Score, and AUC Score to determine the performance of detecting BIC and clean (non-BIC) commits. We extract basic features for all the subject systems from GS, TS, and TP. We have combined those features and got four extra (GS+TS, GS+TP, TS+TP, GS+TS+TP) feature lists. We obtain the results by prioritizing features (eliminating unimportant features using RFE) from all seven feature combinations. To compare the performance of BIC and non-BIC detection without eliminating unimportant features, we also reported the results using all the features from GitHub Statistics (GS-ALL, no feature eliminated) for all the subject systems. Based on the results of this investigation, we are answering our research questions as follows. 

\subsection{Answer to the \textbf{RQ1}}
\textbf{How can we determine whether developers’ coding syntax patterns could be responsible for inducing bugs in a software system?}

This research question was the most important motivation for doing this study. We wanted to see whether the coding syntax style could induce buggy or faulty code fragments in the software system. The detailed extraction and encoding process of TS and TP features discussed in Sections \ref{detailTS-TP}, and \ref{encoding-ts-tp} explain the different steps of identifying the coding syntax styles/ patterns and encoding the labeled commits with those identified TS and TP values. 
For instance, if a different developer writes the test.java file demonstrated in Figure \ref{fig:SourceFeatureExtractionSrcML}, it could have a different coding structure, which will lead to obtaining a completely different list of TS and TP feature values. As the source code patterns encoded by TS and TP values improve BIC detection performance in all eight software projects compared to the conventional feature values (GS), we can say that these patterns might also be responsible for inducing bugs in software systems.  Evaluation of the results of this study verifies this assumption as follows.  

\begin{figure}
\includegraphics[width=\textwidth] {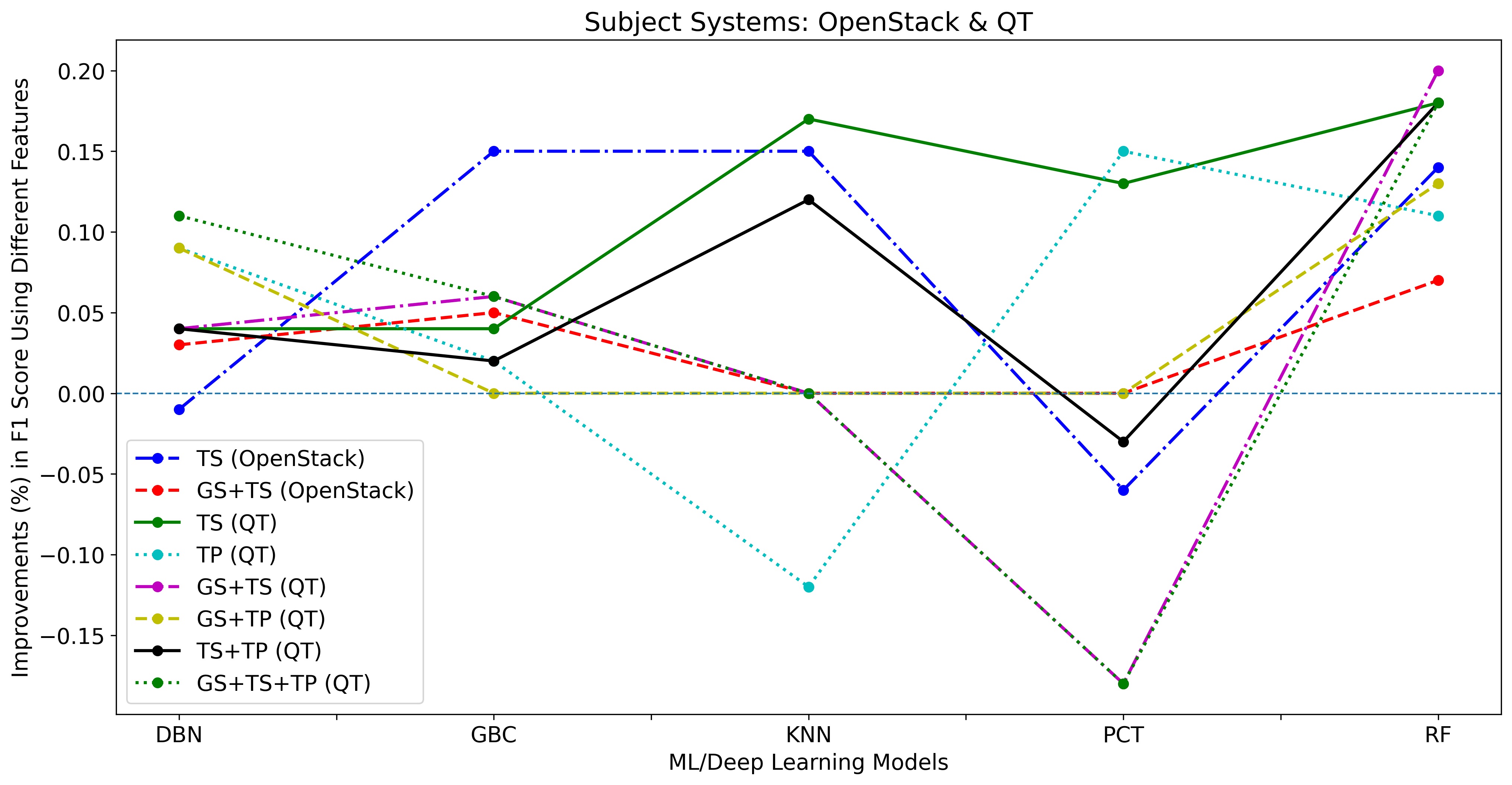}
\caption{We are comparing the improvements of F1~Scores using different Feature Combinations using five different machine learning-based classifiers to predict buggy and non-buggy commits from two popular datasets (OpenStack and QT). Here, the horizontal line at 0 indicates the baseline F1~Scores, and the lines and dots above and below the horizontal line indicate the percentage of improvements and decreases in F1~Scores, respectively, when we use different combinations of our proposed features.}
\label{fig:fscoresQTOS}
\end{figure}

\begin{figure}
\centering
\includegraphics[width=\textwidth] {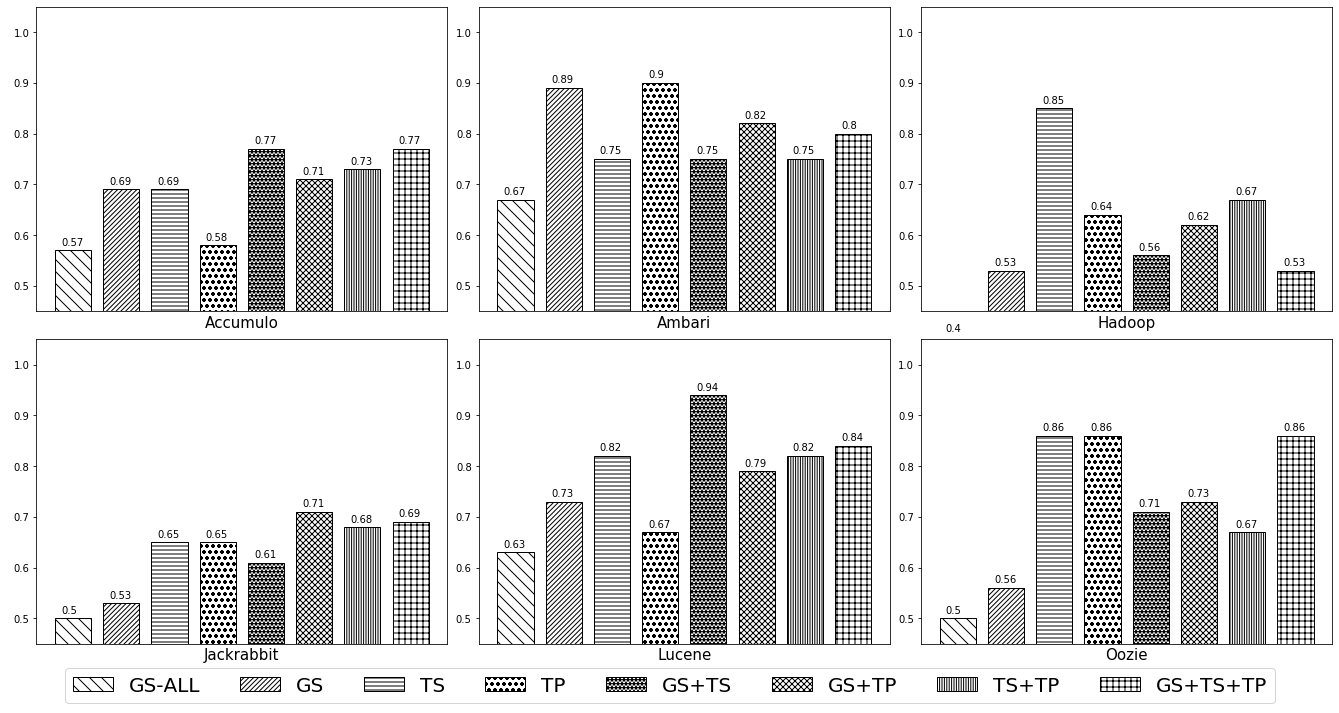}
\caption{Performance Comparison by Precision in different subject systems using different feature combinations.}
\label{fig:bicFEngPrecision}
\end{figure}

\begin{figure}
\centering
\includegraphics[width=\textwidth] {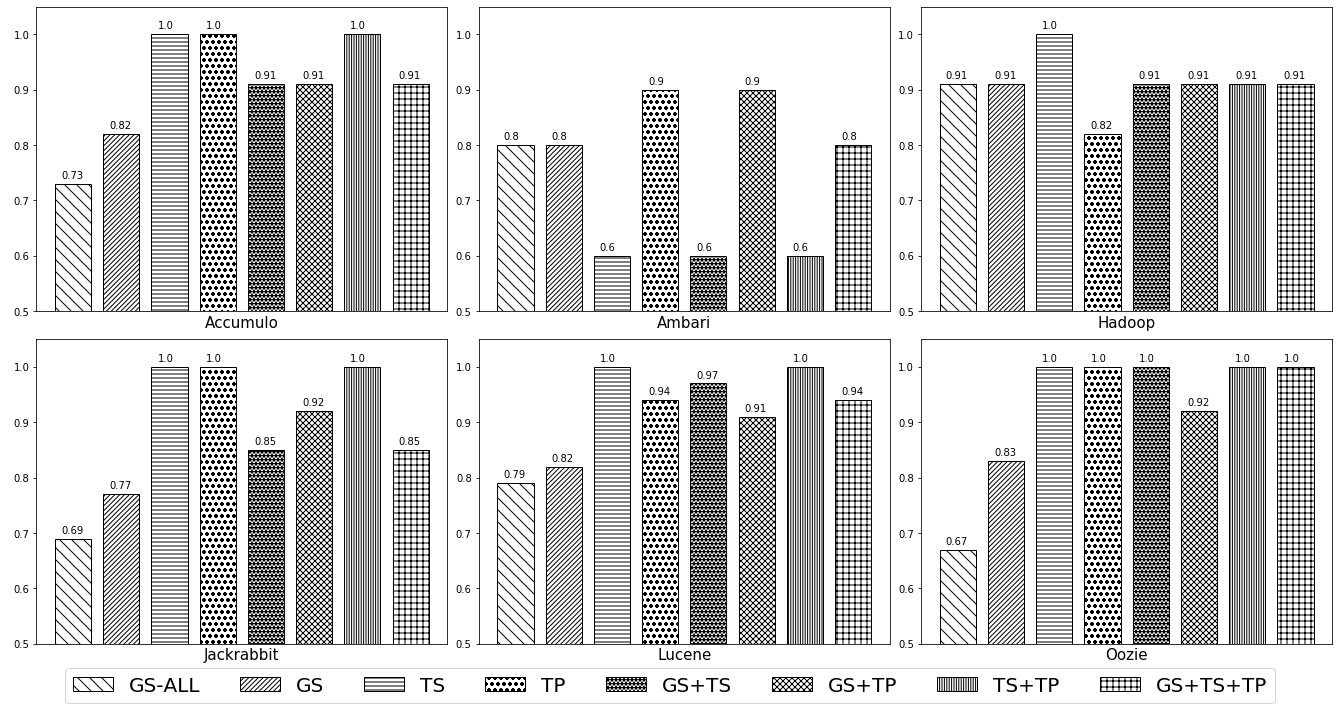}
\caption{Performance Comparison by Recall in different subject systems using different feature combinations.}
\label{fig:bicFEngRecall}
\end{figure}

\begin{figure}
\includegraphics[width=\textwidth] {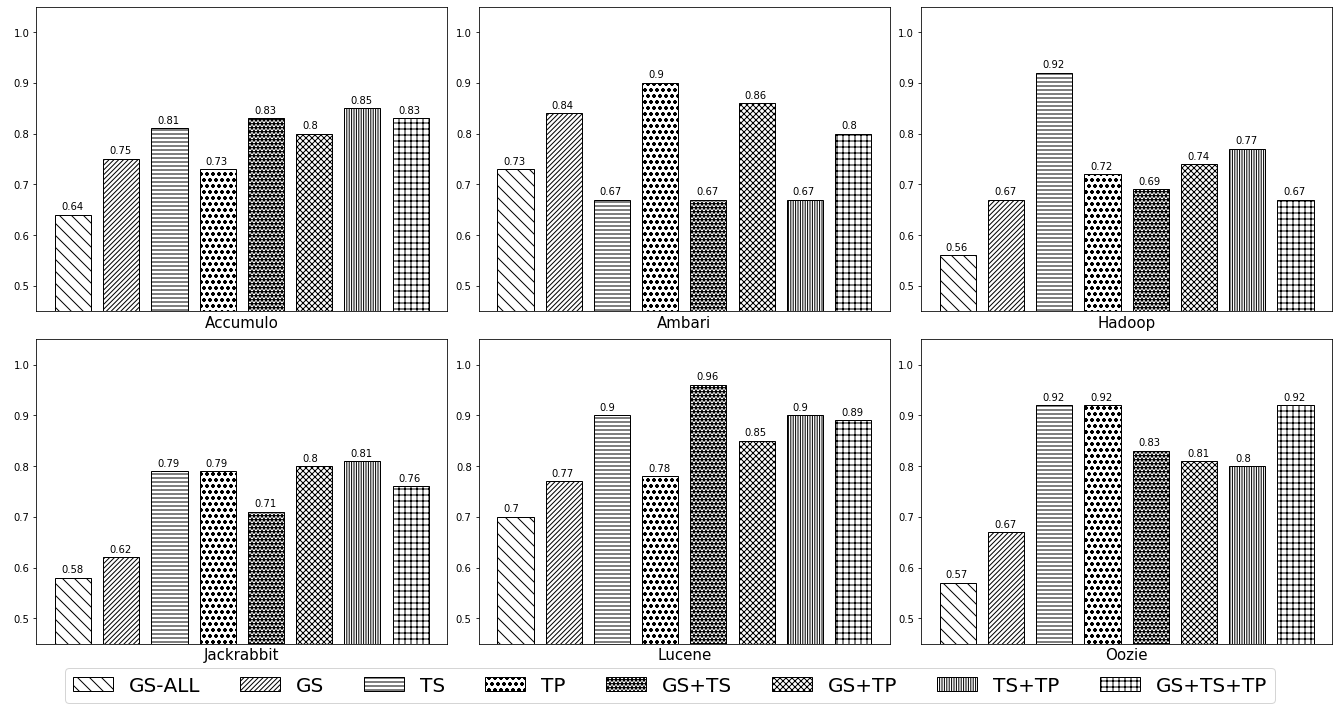}
\caption{Performance Comparison by F1~Score in different subject systems using different feature combinations.}
\label{fig:bicFEngF1Score}
\end{figure}

We show our results (Precision, Recall, and F1~Score) from the six manually labeled subject systems in Table \ref{tab:data-summary-i} in Figures \ref{fig:bicFEngPrecision}, \ref{fig:bicFEngRecall}, and \ref{fig:bicFEngF1Score} and two automatically labeled subject systems in Table \ref{tab:data-summary-ii} in Figure \ref{fig:fscoresQTOS}. These figures clearly show that we can detect BIC (and the non-BIC) in all software systems using only the TS or TP feature with higher performance measures (F1~scores) than the most commonly used GS-ALL and GS features. As ML-based detection models can distinguish BIC and non-BIC using the feature values representing the developer's coding syntax style/ pattern, we can provide an affirmative answer to this research question. However, we can also argue that the ML-based model mainly utilizes those specific coding syntax styles/patterns to detect BIC from the software commits. Therefore, these coding syntax patterns must be dealt with very carefully while doing change operations in the codebase. 

The results obtained using the six subject systems in Table \ref{tab:data-summary-i} show that considering all the evaluation metrics (Precision, Recall, F1 Score, and AUC scores in all the subject systems), the performance obtained by only GS features is lower than the performance obtained by the use of TS and TP features. Besides, using all the 12 feature values (GS-ALL) extracted from GS has a lower performance than the prioritized features from GS, where we removed some unimportant features in all the subject systems. Comparing the BIC detection performance between GS-ALL and prioritized GS indicates that if we take only important features from GS-ALL, it will improve the performance, but this is not enough to detect all the BICs. On the other hand, using feature values extracted from TS and TP can improve the detection performance at a statistically significant level. This scenario provides the answer to this research question. In all the subject systems, either TS or TP or both features improved performance more than the GS-ALL (non-prioritized) and GS (prioritized) features.

The results of two automatically labeled subject systems in Table \ref{tab:data-summary-ii} also provide improved F1~Scores using all five different ML-based classification models in Figure \ref{fig:fscoresQTOS}. The horizontal line at 0.00 demonstrates the F1~Score using GS features only, and the points above and below the 0.00 line indicate the percentage (\%) of improvements and decrease in F1~Scores, respectively. We show the gain (in \%) in F1~Scores using all the different feature combinations of GS, TS, and TP. Our results show inclusion of TS and TP features improves the F1~Score in all five ML-based classification models. Some feature combinations in PCT and KNN provide a decrease in F1~Scores, but these are only six cases out of 40 total test cases in two subject systems.s

The distribution of the AUC scores shown in Figure \ref{fig:boxAUCscores} also supports our findings. Most of the AUC scores obtained using TS and TP features are higher than GS-ALL and GS features. We find the highest distribution of AUC scores using TS and GS+TS feature values. TP features also provide higher AUC scores but could not outperform GS and GS+TS. Although other feature combinations such as GS+TP, TS+TP, and GS+TS+TP also provide improved AUC scores compared to GS-ALL and GS features, it gives a similar distribution of AUC scores to other TS and TP feature combinations. 

\begin{figure}
\includegraphics[width=\textwidth] {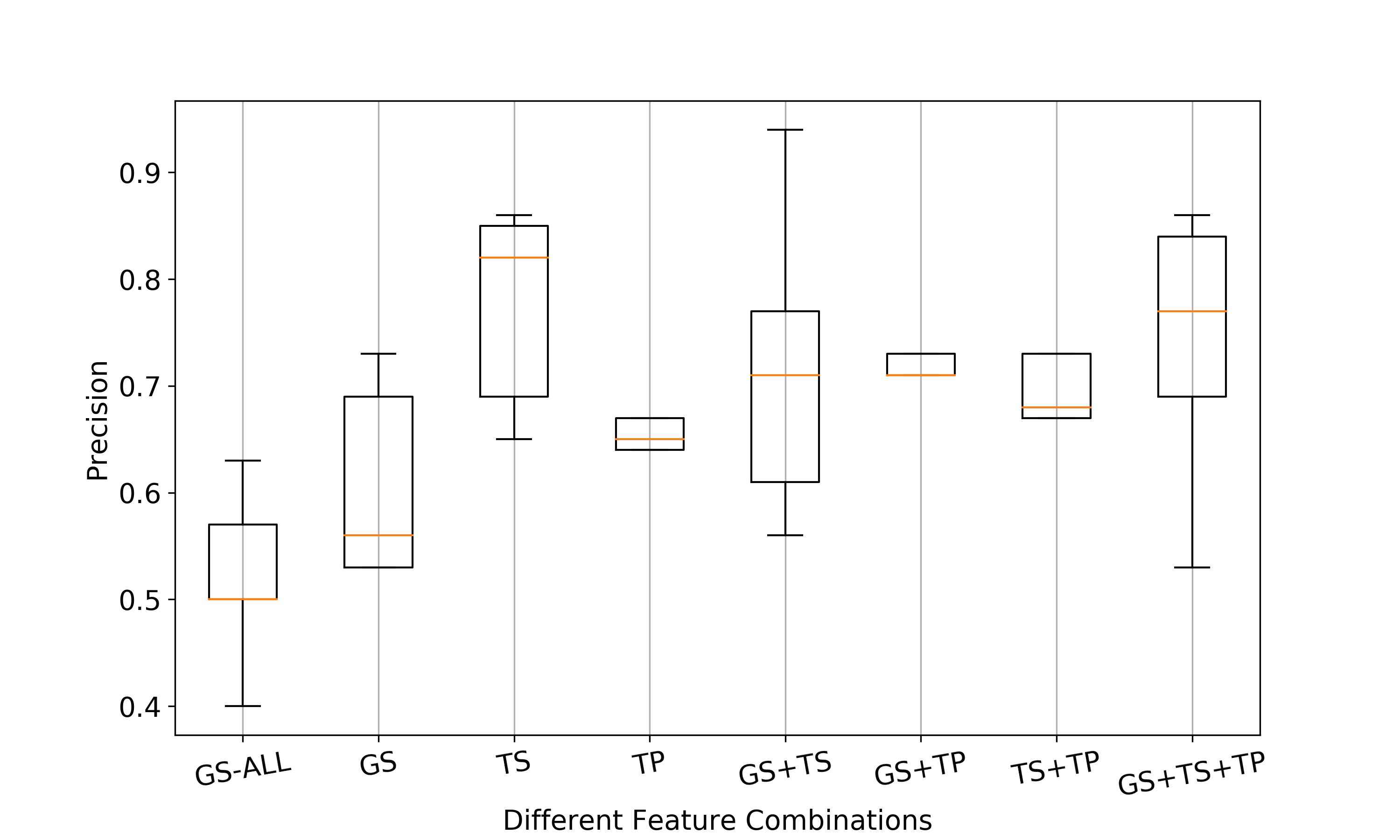}
\caption{Distribution of precision shows TS features provide much higher scores than other features.}
\label{fig:boxPrecisions}
\end{figure}

\begin{figure}
\includegraphics[width=\textwidth] {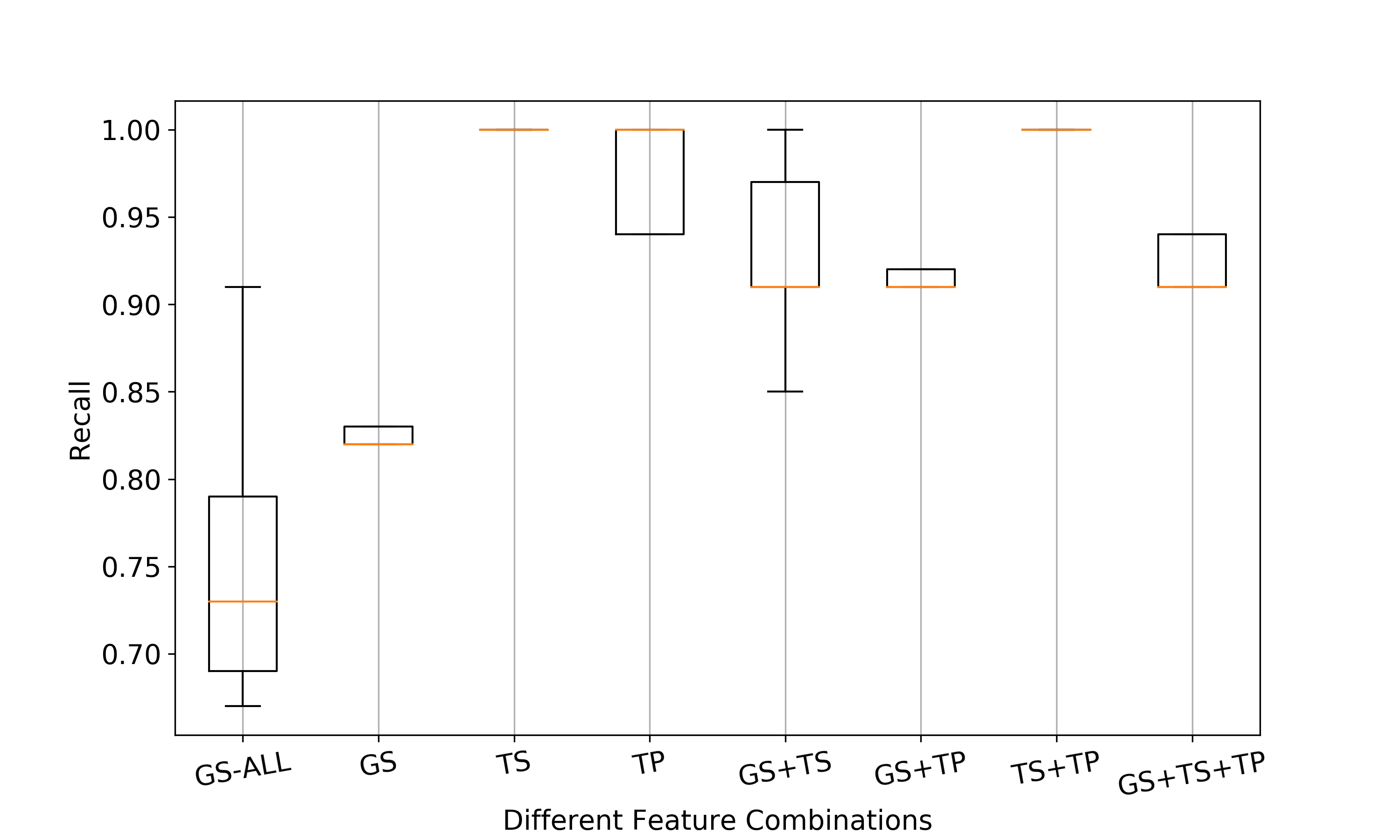}
\caption{Distribution of recall shows the addition of any of the TS or TP or both provides higher scores than GS and GS-ALL alone.}
\label{fig:boxRecalls}
\end{figure}

\begin{figure}
\includegraphics[width=\textwidth] {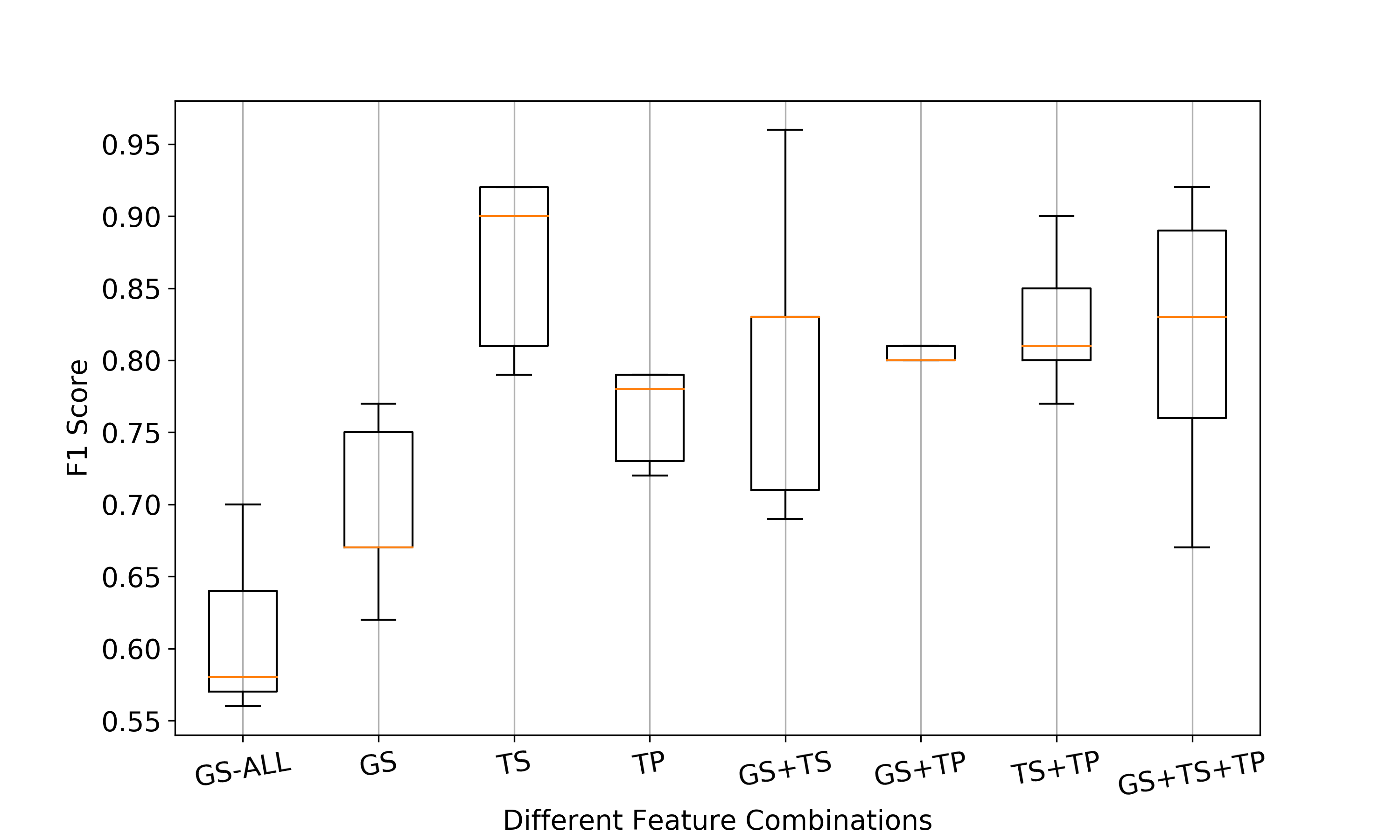}
\caption{Distribution of F1~Scores shows the addition of any of the TS or TP or both provide higher scores than GS and GS-ALL alone.}
\label{fig:boxFscores}
\end{figure}

\subsection{Answer to the \textbf{RQ2}}
\textbf{Do the features extracted from developers' coding syntax patterns provide significantly better performance compared to the other feature values?}

We performed the Wilcoxon Signed Rank Test to see if TS and TP-based feature values significantly improved the performance parameters (e.g., Precision, Recall, F1 Score). This statistical significance test will compare the results using TS and TP feature values with those obtained using feature values from GitHub Statistics (e.g., GS-ALL, GS ). For instance, we would like to examine whether the Precision/ Recall/ F1 Score obtained by GS+TP features is significantly better than the results obtained by GS features. Thus, we applied both GS+TP and GS features independently in our six subject systems and obtained Precision, Recall, and F1 Scores from each implementation, as shown in Table \ref{tab:demo-wilcoxon-rank-test}. Therefore, we got six pairs of observations for each of the Precision, Recall, and F1 Score, and we used them to calculate the differences between each observation pair. Then, we used the differences between these observation pairs to perform the Wilcoxon Signed Rank Test utilizing the SciPy library \cite{SciPy-NMeth2020} available in the Python programming language. Finally, we performed a significance test for each Precision, Recall, and F1 Score separately.

\begin{figure}
\includegraphics[width=\textwidth] {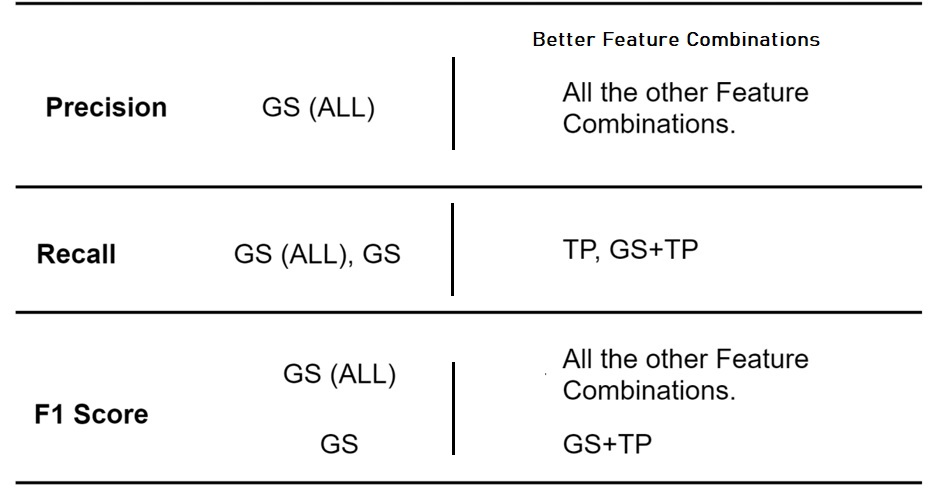}
\caption{Result of The Wilcoxon Signed Rank Test ($p<0.05$). The better feature combination on the right means we obtain statistically-significant improved results when we use these features compared to the ones on the left.}
\label{fig:WilcoxonRankTest}
\end{figure}

\begin{figure}
\includegraphics[width=\textwidth] {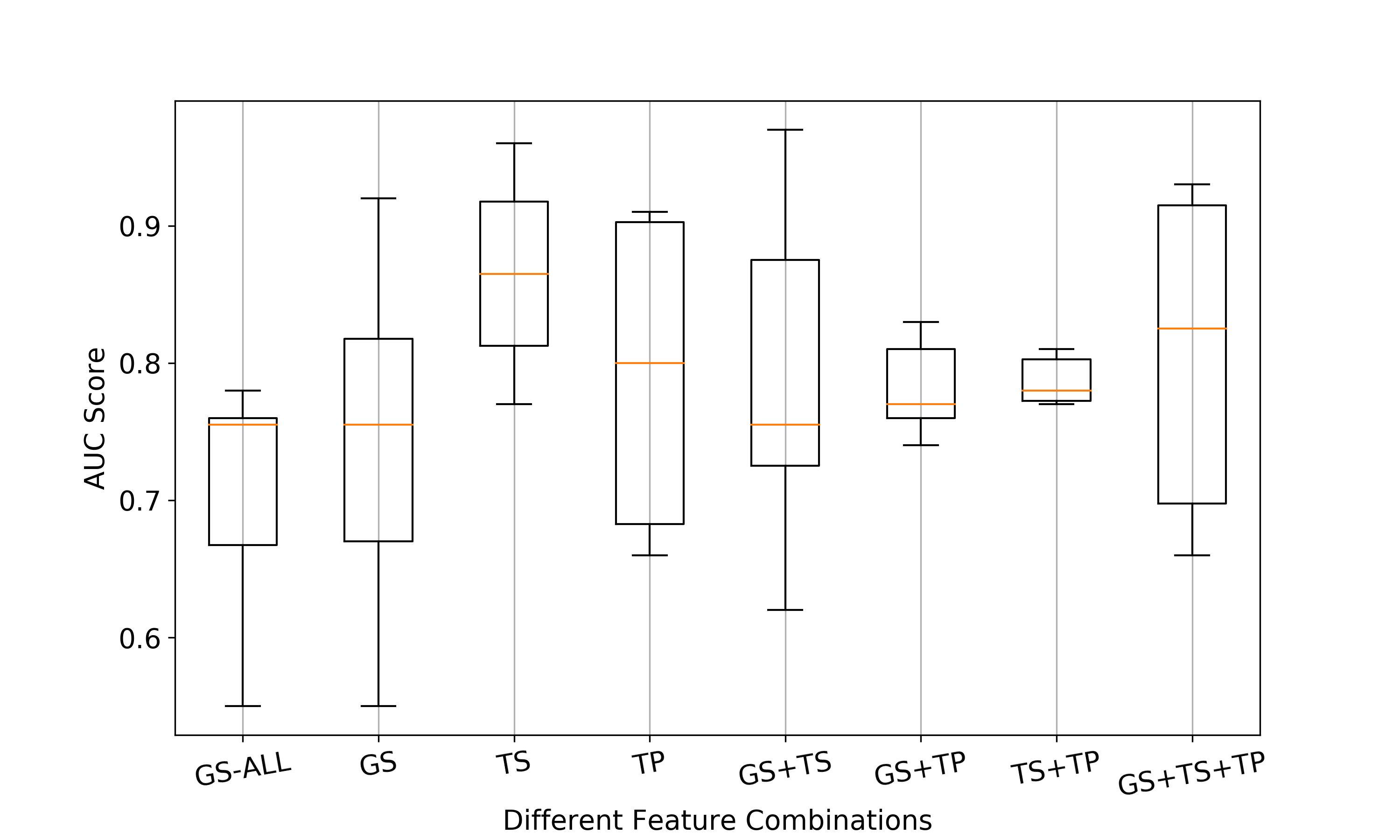}
\caption{Distribution of AUC scores using different feature combinations.}
\label{fig:boxAUCscores}
\end{figure}

A summary of the results obtained from the significance test of the features is given in Figure \ref{fig:WilcoxonRankTest}, which shows the results of a feature list are significantly different from the results of corresponding features using a $\vert$ (vertical bar) symbols. To compare which result is better, we can refer to the distribution of the results shown in Figures \ref{fig:boxPrecisions}, \ref{fig:boxRecalls}, \ref{fig:boxFscores}, and \ref{fig:boxAUCscores}. When we consider the precision of BIC detection, we can see that prioritized features from GS, TS, TP and any combinations of these features perform significantly better than using all the 12 features from GitHub Statistics (GS-ALL). TP features alone or a combination of TP and GS, providing significantly better results than GS-ALL and GS if we consider the recall results. The significance test of the F1~Score also shows any other feature combinations are significantly better than the GS-ALL features and GS+TP are better than GS. Therefore, we can conclude that features extracted from TP in all the scenarios provide significantly better results when combined with GS than the GS-All or GS features alone. Thus, our findings in this study provide enough evidence to believe that more source code-related features (TS, TP) can increase the detection accuracy significantly in identifying BIC using machine learning models. 

\begin{table}
\centering
\caption{\textsc{Preparing set of Observations for The Wilcoxon Signed Rank Test from Our Results}}
\label{tab:demo-wilcoxon-rank-test}
\begin{tabular}{|c|c|c|c|c|c|c|} 
\hline
\multirow{2}{*}{\begin{tabular}[c]{@{}c@{}}\textbf{Performance }\\\textbf{ Measure}\end{tabular}} & \multicolumn{6}{c|}{\textbf{Subject Systems}}                                                   \\ 
\cline{2-7}
                                                                                                  & \textbf{S1}   & \textbf{S2}    & \textbf{S3}   & \textbf{S4}   & \textbf{S5}   & \textbf{S6}    \\ 
\hline
\textbf{Precision (GS)}                                                                           & 0.69          & 0.89           & 0.53          & 0.53          & 0.73          & 0.56           \\ 
\hline
\textbf{Precision (GS+TP)}                                                                        & 0.71          & 0.82           & 0.62          & 0.71          & 0.79          & 0.73           \\ 
\hline
\textbf{Difference}                                                                               & \textbf{0.02} & \textbf{-0.07} & \textbf{0.09} & \textbf{0.18} & \textbf{0.06} & \textbf{0.17}  \\ 
\hline
\multicolumn{7}{|l|}{}                                                                                                                                                                              \\ 
\hline
\textbf{Recall (GS)}                                                                              & 0.82          & 0.80           & 0.91          & 0.77          & 0.82          & 0.83           \\ 
\hline
\textbf{Recall (GS+TP)}                                                                           & 0.91          & 0.90           & 0.91          & 0.92          & 0.91          & 0.92           \\ 
\hline
\textbf{Difference}                                                                               & \textbf{0.09} & \textbf{0.10}  & \textbf{0.00} & \textbf{0.15} & \textbf{0.09} & \textbf{0.09}  \\ 
\hline
\multicolumn{7}{|l|}{}                                                                                                                                                                              \\ 
\hline
\textbf{F1 Score (GS)}                                                                            & 0.75          & 0.84           & 0.67          & 0.62          & 0.77          & 0.67           \\ 
\hline
\textbf{F1 Score (GS+TP)}                                                                         & 0.80          & 0.86           & 0.74          & 0.80          & 0.85          & 0.81           \\ 
\hline
\textbf{Difference}                                                                               & \textbf{0.05} & \textbf{0.02}  & \textbf{0.07} & \textbf{0.18} & \textbf{0.08} & \textbf{0.14}  \\
\hline
\end{tabular}
\end{table}

\clearpage
\subsection{Answer to the \textbf{RQ3}}
\textbf{How generalized are the extracted features from one software system to the others?}
\subsubsection{Manually Labeled Datasets of Table \ref{tab:data-summary-i}}
\begin{figure}
\includegraphics[width=\textwidth] {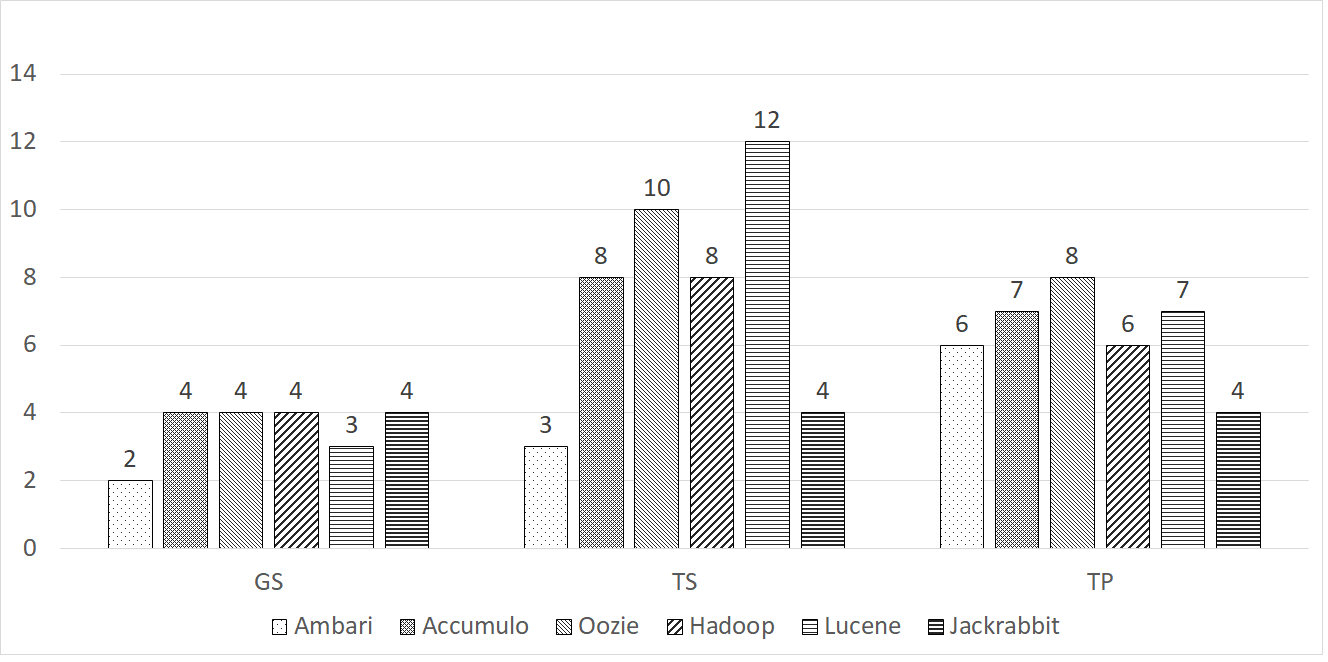}
\caption{Size of Best Feature Lists for Detecting BIC from each of the Subject Systems}
\label{fig:sizeBestFSets}
\end{figure}

We investigated the number of features in the best feature lists for each GS, TS, and TP feature in all the subject systems. We represented the list sizes in different subject systems according to the number of samples of the subject systems in Figure \ref{fig:sizeBestFSets}. In our investigation, Ambari has the smallest number of data samples (Table \ref{tab:data-summary-i}), Accumulo is next to Ambari, and Jackrabbit has the maximum number of data samples. By sorting the number of features in the list of best features, we wanted to see whether the best feature list depends on the number of available data instances of a subject system. We show the number of available data instances from each subject system in Table \ref{tab:data-summary-i}. 

Figure \ref{fig:sizeBestFSets} shows that the number of features to provide the best result increases for most subject systems. In this figure, the subject systems are arranged based on the number of available data instances. Thus, the left-most subject system (e.g., Ambari) contains the least number of data instances, and the right-most subject system (e.g., Jackrabbit) has the highest number of data instances.  Although Hadoop, Lucene, and Jackrabbit show a decline while using different features (GS, TS, TP), all the other three subject systems show an increase in this number with the rise in data samples in the subject systems. To verify this assumption, we conducted both Pearson’s \cite{PearsonCorrelation} and Spearman's \cite{SpearmanCorrelation} correlation tests and found that the number of GS features is positively correlated, and TS and TP features are negatively correlated with the number of data instances in software projects under investigation for providing the best BIC detection performance, and these correlations are not statistically significant. Therefore, we can say that the number of features is not directly related to the number of data instances under investigation. There might be unique feature sets for each software project that may provide better BIC detection performance using different ML models. 

We also investigated whether a set of features is generalizable from one subject system to another. Analyzing the list of best features, we can see that for almost all the subject systems, even though the number of features in different best feature lists are the same, the feature identifiers are different. For example, though there are 12 features available in the case of GS features, only 3 to 4 features can give the best BIC detection performance, and they contain at least one or more distinct features among various available subject systems. Similarly, different sets containing only 6 to 12 features can provide the best result from the thousands of features when considering the TS and TP-based features.

These findings emphasize the importance of selecting an appropriate number of most relevant features specialized for each subject system to identify BIC based on the data instance size of the software project available to train and test a machine learning model. A fixed set of features might not provide the best results for all the software projects to identify bug-inducing software commits using ML models.  

\subsubsection{Automatically Labeled Datasets of Table \ref{tab:data-summary-ii}}

Our investigation evaluates whether source code syntax pattern (TP) and sequence (TS) based features are generalizable toward different software projects and ML-based classification models. We apply BIC detection using five classification models on the automatically labeled datasets (OpenStack and QT) of Table \ref{tab:data-summary-ii}. Our results are in Figure \ref{fig:fscoresQTOS}, where we compare the percent (\%) of improvement in F1~Scores of BIC detection using our extracted features with GS features. Our results show that in all five classification models and two subject systems, our proposed features improve the F1~Scores. The highest improvement (20\%) in the F1~Score was obtained using Random Forest (RF) classifier in the subject system QT using the GS+TP features. All the other feature combinations also improved F1~Scores compared to the GS feature using RF classifier in both the OpenStack and QT subject systems. Although in some cases, Perceptron (PCT) and K-Nearest Neighbour (KNN) algorithms show a decrease in F1 Scores in both the subject systems, they still show improvement in F1 scores with other feature combinations. Therefore, we can conclude that in most cases, our proposed features are generalizable toward different ML-based classification models and software projects to detect BIC in higher F1 Scores than the GS features. 

\vspace{1cm}
\subsection{Answer to the \textbf{RQ4}}
\textbf{Do the features extracted from developers’ coding syntax patterns enhance the explainability of BIC detection from software systems?}

\citet{CitePyExplainer} published a state-of-the-art tool named PyExplainer to explain the underlying condition for machine learning predictions. We use their tool to explain our BIC detection using the conventional (GS) and our proposed Token Pattern (TP) features. The results of the most frequent conditions from PyExplainer in the QT subject system are shown in Table \ref{tab:demo-pyexplainer}. We show the top five most frequent feature conditions from GS and TP features that detected buggy commits using Random Forest (RF) algorithm in the table, and all the other conditions are publicly available to access in the GitHub repository of our investigation. In the table, we can see that it is very difficult to make proper reasoning for buggy commit detection considering GS features from the identified conditions by PyExplainer. For example, the top five most frequent conditions obtained from the result of PyExplainer are some range of values of some GS features such as Awareness, Age, Line Added, and Number of functions, but it does not show any specific reasoning that what a developer can do to avoid bug proneness in a software project for a certain given condition. A software development environment with a set of developers may have a fixed set of given conditions, such as developer awareness of the software project, age of the developer, number of lines to be added or changed, etc. These given conditions might be difficult to modify instantly or within a short period.  Therefore, if we detect buggy software commits using those unchangeable conditions as feature values, it can provide enough meaningful or explainable reasoning to the software developer about how they can avoid the bug proneness of a particular change.  

\begin{table}
\caption{\textsc{Comparing Key Conditions to Detect Commit Bug Proneness by PyExplainer}}
\label{tab:demo-pyexplainer}
\centering
\begin{tabular}{|l|l|} 
\hline
\textbf{Feature Names}                                                                                                                & \textbf{PyExplainer Condition}                                              \\ 
\hline \hline

\multicolumn{2}{|l|}{\textbf{i. Code Churn Based Features}}                                                                                                                                     \\ 
\hline
Awareness$^1$                                                                                                & 0.0050 $<$ asawr $<=$ 0.0250                                             \\ 
\hline
Awareness$^1$                                                                                                & 0.1149 $<$  osawr $<=$ 0.1150                                             \\ 
\hline
Age$^2$                                                                                                                 & 694101.53 $<$  age $<=$ 647989.44                                         \\ 
\hline
Line Added$^1$                                                                                                          & 57.59 $<$ la $<=$ 621.30                                                 \\ 
\hline
Number of Functions$^3$                                                                                                & 4.14 $<$ nf $<=$ 2.26                                                    \\ 
\hline \hline

\multicolumn{2}{|l|}{\textbf{ii. Code Syntax Pattern (TP) Based Features}}                                                                                                                       \\ 
\hline
Declaring a Variable Name                                                                                               & 1.84 $<$ decl\_name $<=$ 15.32                                           \\ 
\hline
\begin{tabular}[c]{@{}l@{}}Expression with Operators \\in IF Condition\end{tabular} & 0.40 $<$ if\_condition\_expr\_operator $<=$ 7.81                         \\ 
\hline
Expression with Operators                                                                                               & 0.47 $<$ expr\_operator $<=$ 2.42                                        \\ 
\hline
\begin{tabular}[c]{@{}l@{}}Expression with Function Call \\with Argument and Expression \\with Variable Name\end{tabular} & 0.375 $<$ expr\_call\_argument\_expr\_name $<=$ 7.20                      \\ 
\hline
Expression with Operators                                                                                               & 0.47 $<$ expr\_operator $<=$ 24.27                                       \\
\hline \hline

\multicolumn{2}{|l|}{\begin{tabular}[c]{@{}l@{}}1.~Does Not Provide Enough Information About Why The Bug Occurred\\2.~Unrealistic Condition\\3.~Contradictory Condition\end{tabular}}    \\ 
\hline
\multicolumn{2}{l}{}                                                
\end{tabular}
\end{table}

On the other hand, if we see the result from PyExplainer using Token Pattern (TP) features for the top five most frequent conditions for detecting buggy commits; we can see that most of the commits are detected buggy because of the presence of specific source code syntax patterns (i.e., number of declared variable names, expression with operators in IF-conditions, different level of nested expression and function call, etc. ). In any given condition of a software development environment, these conditions of a source code syntax can easily be updated to avoid or minimize bug proneness of a software project. These scenarios provide evidence to argue that, Token Patterns (TP) as feature values in the ML-based classification model enhance the explainability of BIC detection from software systems compared to the GS features. We plan to do more investigation about this scenario with some other subject systems in the future.

\section{Threats to Validity}
\label{threats-validity}
\subsection{\textbf{Subject Systems \& Programming Languages}}
Subject systems in this study are written in Java, CPP, and Python programming languages. Therefore, we can argue that the results of this study may not be generalizable to the subject systems of other programming languages not used in this investigation. However, the subject systems used in this investigation are widely used in similar studies and are of diverse varieties. Thus we are confident about the outcome. Furthermore, since we wanted to compare our newly introduced features (TS, TP) to the most common features (GS) by exploiting the manually and automatically labeled data of eight of those systems, we used these widely used diverse varieties of Java, CPP, and Python systems. Regarding systems of other programming languages, we note the technique used may provide similar results for similar programming paradigms. Of course, further investigation is necessary, which remains our future work.  

\subsection{\textbf{Dataset Labeling for Buggy and Non-buggy Commits}}
\citet{2019:FSE:Wen:EEC:3338906.3338962} reported that automatically labeling buggy and non-buggy commits leave many incorrect labeling. On the other hand, getting enough data instances to apply ML classification models in manually labeled buggy and non-buggy software projects is difficult. For example, only 642 data instances of the buggy and non-buggy commits are available in six software projects from the manually labeled dataset \cite{2019:FSE:Wen:EEC:3338906.3338962} used in this study. A small number of data instances or incorrectly labeled data instances both have a detrimental effect on the generalizability of results obtained by ML models. Therefore, we used the six manually, and two automatically labeled software projects of buggy and non-buggy commit to performing our investigation. We accomplish this investigation in two steps. First, we apply our technique in manually labeled datasets of the buggy and non-buggy commits by \citet{2019:FSE:Wen:EEC:3338906.3338962} and show that our proposed features improve the performance of the ML model to detect buggy commits. We then apply five different ML classification models in two automatically labeled datasets having 4,014 data instances. In both experimental setups, we show that using our proposed feature values improves the performance of ML models to detect buggy commits. Our goal is not to contradict the results of any state-of-the-art JIT defect prediction method. Instead, we applied five different machine-learning models and eight different feature combinations. We showed in almost all the ML algorithms that using our proposed features improves the detection performance measures of bug-inducing commits. Therefore, we believe selecting any other dataset in this investigation should have similar findings. However, in the future, we want to extend this study using other software projects of different programming languages to substantiate the conclusion. 

\subsection{\textbf{Machine Learning Classification Models}}
We also tried applying K-Nearest Neighbour (KNN) and Logistic Regression to detect BIC from the manually labeled datasets in Table \ref{tab:data-summary-i}. However, as these datasets contain a smaller number of data instances than a large number of features (different combinations of GS, TS, and TP), KNN and LR failed to accurately estimate the precision and recall of different implementations in detecting BIC. Therefore, we only used the Random Forest (RF) classifier to identify BIC from the manually labeled datasets and utilized four more ML classification models in the automatically labeled datasets in Table \ref{tab:data-summary-ii}. As different ML algorithms work differently, it may be questionable whether our results obtained using the manually labeled datasets are generalizable for different ML algorithms or not. \citet{fault-prediction-study-survey} reported that ML models' performance mainly depends on the nature of data, the quality of features, and proper parameter tuning. This study's main goal was to compare the ML-based model performance by using the features GS, TS, TP, and their different combinations. We also applied five different classification models on the larger datasets shown in Table \ref{tab:data-summary-ii}. The results in figure \ref{fig:fscoresQTOS} show that in most cases, the use of software TS and TP (source code syntax pattern ) based features improve the BIC detection performance in all the ML models. If a different ML algorithm is used, it should equally affect all the feature combinations. Therefore, the comparison scenario obtained in this study should remain the same in different ML-based detection models. Although, more investigations using more massive datasets and other ML models are required to verify this scenario, which we will do in future studies.  

\section{Related Work}
\label{related-work}
Studies related to detecting or predicting Bug Inducing Commit(s) attracted the attention of researchers due to their massive impact on software systems. Some studies \cite{Kim:Classify:Clean:Buggy, Sliwerski-2005-induce-fixes-journal, Rel:Kim:2006:AIB:1169218.1169308, Eyolfson:2011:TDD:1985441.1985464} tried to find the changes in a software system that is responsible for the first introduction of a bug, while some other studies \cite{Asaduzzaman:Bug:Inducing:USASK, Bavota:2012:RIB:2477171.2477400, Bernardi:Developer:Induce:Bug, Canfora:How:Long:Bug:Survive} try to identify the bug fixing commits and then link it to its corresponding bug-inducing  commit using SZZ Algorithm \cite{Sliwerski-2005-induce-fixes-journal}, and its variants \cite{Framework:SZZ, Text:Based:Bug:Inducing, Rel:Kim:2006:AIB:1169218.1169308}. There are many studies to understand the essential characteristics of bug-inducing  commits using machine learning models \cite{Asaduzzaman:Bug:Inducing:USASK, Bavota:2012:RIB:2477171.2477400, Bernardi:Developer:Induce:Bug, Canfora:How:Long:Bug:Survive, Ell:Failure:Inducing:Developer:Pair, Eyolfson:2011:TDD:1985441.1985464, Kim:2006:LDT:1137983.1138027, Sliwerski:2005:HRR:1095430.1081725}. \citet{Rel:Aversano:2007:LBC:1294948.1294954} did a study on two relatively small Java systems and applied five machine learning algorithms with 10-fold cross-validation to predict whether a change is likely to be buggy or not. They only used the features extracted from the GitHub repository. \citet{Rel:Fukushima:2014:ESJ:2597073.2597075} and \citet{Kamei:2013:LES:2498737.2498844} worked for predicting defects in software systems, but none of them showed any use of code quality measures in their prediction models. There are also some other studies \cite{Kim:Classify:Clean:Buggy, Rel:Kim:2007:PFC:1248820.1248881, Rel:RePEc:ids:ijrsaf:v:7:y:2013:i:1:p:17-31, Rel:2013:Shivaji:Reducing:Feature:Bug:Prediction, Rel:2015:Yang:Deep:JIT:Defect:Prediction, Rel:Kim:2006:AIB:1169218.1169308, Sliwerski-2005-induce-fixes-journal, Rel:Wen:2016:LLB:2970276.2970359} for predicting the likelihood of being a commit bug-inducing, but no earlier studies explored the impact of code quality measures or developers' patterns of source code in this research domain. We tried to minimize the gap by proposing and testing a method to encode software commits using developers' coding syntax patterns while doing software revisions.  Our study compares the results to detect bug-inducing commits using the conventional features of the existing methods and our proposed developers' source code pattern-based features. 

 \citet{Rel:2013:Shivaji:Reducing:Feature:Bug:Prediction} applied various feature selection techniques that are commonly used in classification-based bug prediction techniques. The techniques drop unnecessary features until optimal classification performance is achieved. The total number of features used for training is extensively degraded, often to fewer than 10 percent of the initial. We also applied a similar technique using the Recursive Feature Elimination (RFE) of the SciKit Learn \cite{scikit-learn} python library for reducing the required number of features to obtain the best BIC detection result.

 \citet{Rel:2018:Wu:ChangeLocator:CLC:3180155.3182516} applied Logistic Regression, Decision Trees, Naive Bayes, and Bayesian Network (BayesNet) to find the changes which induce crush from several subject systems. Our work complements the work done by \citet{Rel:2013:Shivaji:Reducing:Feature:Bug:Prediction} in several ways. Although they added some features from source code metrics and BOW+ implementation on the source code, they did not focus on developers' coding syntax patterns. Besides, they did not normalize the source code by tokenization \cite{tokenization} to find the optimal feature values. Our study tried to overcome both limitations by extracting features from the tokenized XML representation of source code fragments.
 
 Our work also complements prior similar studies in several ways. First, most of the prior studies tried to detect the likelihood of a commit to bug-inducing or clean (non-bug-inducing) using only the statistical measures extracted from the GitHub repositories of the respective subject systems. We added two different types of extra features (Token Sequence-TS and Token Pattern-TP) by processing the source code of commit patches. The addition of source code pattern-based features provides the likelihood of a commit being buggy or not, and it also provides an idea of which types of source code patterns are more likely to induce bugs in software systems. We also executed the BIC detection technique without those features and compared the effect of adding our features. Our comparisons show we improved the performance of BIC detection with the addition of source code TS and TP-based features. Second, we applied the Recursive Feature Elimination (RFE) algorithm to investigate the importance of each feature in an ML-based BIC detection model. Third, we investigated an approach to find the best features by starting with the top feature and adding only those features to the list, improving detection performance. After completing the full iteration with all the features, we got a relatively small list of features providing better BIC detection results. We believe these additions made this study much more unique compared to the other related works. 
 
 This paper is an improved and rewritten version of a chapter from the M.Sc. thesis \cite{nadim-thesis-msc} defended by the first author of this paper in September 2020. The thesis is available online as the official \textbf{Graduate Theses and Dissertations HARVEST} by the University of Saskatchewan, Saskatoon, Canada, which includes two more related studies for detecting and reducing \cite{nadim-iwsc-2020} bug introduction in software systems. Our key focus in these studies is finding new tools and techniques that can improve the detection and prevention of probable bug-inducing commits in software systems compared to the existing related studies. This study is an essential step towards that goal. We proposed two new types of feature values (Token Sequence-TS and Token Pattern-TP), which can play an essential role in detecting bug-inducing  commits using machine learning models. Our proposed TS, TP-based source code encoding, can also open new research directions in source code representation and analysis related to software bug-proneness and bug-fix patterns \cite{CharacterizationRepeatedFixes, MemoriesBugFix, promiseDataset}, which we intend to do in future research. 

\section{Conclusion \& Future Work}
\label{conclusion-future-work}
Our investigation introduced two new types of features, one represents the token sequence of source code (TS), and the other represents the hierarchy of these tokens (Token Pattern-TP), which we utilized as feature values for detecting bug inducing commits (BICs) from eight subject systems. Thus, The feature values of TS and TP represent the developers' coding syntax style/ patterns. We extracted thousands of bug-inducing and bug-fixing source code patterns (TP) and their sequences (TS) from the history of these software projects and applied feature prioritization. We then applied Random Forest Classifier to detect Bug Inducing Commits (BICs) from the six manually labeled open-source Java projects using these TS and TP-based and conventionally used GitHub statistics (GS) based feature values. We also applied five different machine learning based classification models to the two subject systems containing automatically labeled bug inducing and clean commits. Finally, we used the BIC detection results using GS-based feature values as the baseline and compared them with TS and TP-based results to determine the performance improvements in detecting buggy commits from the software systems. Our results and analysis of four research questions show that the features representing developers' coding syntax style/ patterns (TS and TP) can increase the performance of detecting bug-inducing commits (BICs) and explainability of the detected results compared to the conventional features (GS) using ML-based detection models.

We also did a significance test of the obtained results using the Wilcoxon Signed Rank Test and found the increase in F1~scores is statistically significant in manually labeled datasets. Therefore, we can conclude that developers' coding syntax style/ pattern could be crucial for introducing bugs in the software systems. It impacts the software quality and reliability. Therefore, new tools and techniques are required to help software developers avoid risky coding syntax patterns and sequences that have been found to induce bugs in the earlier history of software systems.

Another observation in this study with the manually labeled datasets from the six subject systems in Table \ref{tab:data-summary-i} is that each software system requires unique features to provide the best detection results if the dataset size is not big enough. The size of the best features list is also different. From this observation, we can conclude that the best feature set and their number depend on how the software system is maintained and how many data samples are available. Different developers maintain different software systems. Their coding syntax style is different; their exposed patterns (TP) and sequences (TS) are also different; this might be the key reason for this observation. However, using all the TS and TP features improve BIC detection in both the subject systems whose datasets have a larger number of data instances in Table \ref{tab:data-summary-ii}. Therefore, if we have larger datasets for training the model using TS and TP-based features, we get improved buggy commit detection performance without prioritizing the features.

In our investigation, we use datasets from software projects written in three different programming languages (i.e., Java, CPP, and Python), and in all software projects, our proposed source code syntax-based features provided improved buggy commit detection performance compared to conventionally used GS features. Our features also perform better when we use a deep learning-based feature extraction technique using a deep belief network. We also investigate the explainability of detected buggy commits using the dataset of one software project, QT. We found the inclusion of TP features provides better explainability about the reason for detecting a buggy commit compared to the GS features.  In future studies, we also want to extend the explainability study using different software projects of different programming languages, which could provide important insight into fixing the buggy commits automatically using the information of how a commit is identified as buggy by the machine learning models. Finding the reason for bug-inducing commits in the software systems may also lead to finding and automatically correcting the buggy patterns by utilizing historical similar (cloned) fixes in the codebase. We also plan to exploit these findings in building IDE-based tools and libraries so that developers can deal with them right away in the IDE. As part of this, we will start with an IDE-based clone detection environment \cite{10.1145/2245276.2231970} with different visualization support as clone visualizations \cite{10.1145/1985404.1985425} but attributed with bug-inducing  patterns and integrate different other clone detection tools used in our studies. We also plan to explore bug-inducing commits in the context of exception handling \cite{7832939} or localizing bugs \cite{10.1145/3236024.3236065} and see to what extent we could integrate them into the IDE support. We plan to study the relationships and their insights between bug-inducing patterns and several other related studies, such as bug propagation through code cloning \cite{MONDAL2019110407, 8094424}, bug-proneness and late propagation tendency in clones \cite{MONDAL201841}, replication of bugs in clones and micro clones \cite{Judith:Bug:Replication, Judith:Micro:Regular:Clone, Islam2017l} or with those clones that have high possibilities of containing bugs \cite{7961508}.

\backmatter

\bmhead{Acknowledgments}
This research is supported in part by the Natural Sciences and Engineering Research Council of Canada (NSERC) Discovery grants, and by an NSERC Collaborative Research and Training Experience  (CREATE) grant.

\section*{Declarations}
\begin{itemize}
\item \textbf{Conflict of interest/Competing interests.} The authors declare that they have no conflict of interest.
\item \textbf{Data availability.} The datasets and source files generated during and/or analyzed during this study are available in our GitHub repository (https://github.com/mnadims/bicDetectionSF/) for readers to investigate
and facilitate any replication study.
\end{itemize}
\bibliographystyle{sn-basic}
\bibliography{sn-bibliography}
\end{document}